\documentclass[superscriptaddress,  11pt, preprint]{revtex4-1}
\pdfoutput=1
\usepackage{graphicx}

\usepackage{amsmath}
\usepackage{amssymb}
\usepackage{bm}
\usepackage{ulem}
\usepackage{comment}

\usepackage{booktabs}
\usepackage{float}

\newcommand{\dd}{\mathrm{d}}
\newcommand{\Tr}{\mathrm{tr}}
\newcommand{\R}{\mathbb{R}}
\newcommand{\norm}[1]{\left\lVert #1 \right\rVert}

\newcommand{\Ct}{\mathcal{C}}

\begin{document}

\title{
	Optimal design of higher-order interactions for mutual synchronization among three collectively oscillating networks
}

\author{Sotaro Kuroiwa}
\email{Corresponding author: kuroiwa.s.32f7@m.isct.ac.jp}
\affiliation{Department of Systems and Control Engineering, Institute of Science Tokyo, Tokyo 152-8552, Japan}

\author{Yuzuru Kato}
\email{katoyuzu@fun.ac.jp}
\affiliation{Department of Complex and Intelligent Systems, Future University Hakodate, Hokkaido 041-8655, Japan}

\date{\today}

\begin{abstract}
	Many networked dynamical systems consist of interacting subnetworks, each of which can exhibit collective oscillations.
	Mutual synchronization among such collectively oscillating networks is ubiquitously observed, and its efficient control is an important engineering problem.
	While control of synchronization has been extensively studied for pairwise interactions, many systems also involve higher-order interactions among three or more subnetworks that cannot be decomposed into independent pairwise couplings.
	In this study, we investigate the optimization of higher-order internetwork coupling for mutual synchronization among three collectively oscillating networks.
	Using phase reduction theory for collectively oscillating networks, we reduce the full network dynamics to a two-dimensional system of phase-difference equations, in which the linear stability of the phase-locked state and the rotational property of the phase differences around it are characterized by a higher-order coupling tensor shared by the three networks.
	We then derive the optimal coupling tensors that maximize either the linear stability or the rotational property for a fixed overall strength of the internetwork coupling.
	As a numerical example, we consider three collectively oscillating networks, each consisting of four FitzHugh--Nagumo elements coupled in a tree structure, and demonstrate that, compared with a random coupling tensor of the same overall strength, the two types of optimal coupling yield faster convergence to the phase-locked state and stronger rotational motion of the phase differences around it, respectively.

\end{abstract}

\maketitle

\section{Introduction}
Many natural and engineered systems can be described as networks of interacting dynamical elements, such as power grids~\cite{motter2013spontaneous}, neuronal networks~\cite{buzsaki2006rhythms}, gene-regulatory networks~\cite{goldbeter1997biochemical}, ecological communities~\cite{bascompte2009disentangling}, and socio-technical systems~\cite{vespignani2012modelling}.
In such systems, the collective dynamics strongly depends on the topology of interactions, including small-world~\cite{watts1998collective} and scale-free~\cite{barabasi1999emergence} structures.
Synchronization is one of the most fundamental collective phenomena on networks and is observed in a wide range of physical, biological, and engineering systems~\cite{kuramoto1984chemical,pikovsky2001synchronization,arenas2008synchronization}.
Understanding and controlling synchronization is therefore one of the central problems in the analysis and design of networked dynamical systems.

A standard theoretical framework for analyzing the synchronization of weakly coupled nonlinear oscillators is phase reduction theory~\cite{kuramoto1984chemical,nakao2016phase,pietras2019network}.
When each oscillator possesses a stable limit-cycle solution, its multidimensional dynamics can be reduced to a one-dimensional phase equation, and the effect of weak coupling is described by a phase coupling function that depends only on the phase differences between the oscillators.
The phase reduction framework has also enabled systematic optimization of synchronization.
For example, optimal waveforms of periodic forcing for entrainment~\cite{harada2010optimal,zlotnik2013optimal} and optimal coupling functions for mutual synchronization~\cite{shirasaka2017optimizing,watanabe2019optimization} have been derived under constraints on the intensity of the forcing or coupling, and periodic waveforms designed within this framework have been used to experimentally realize prescribed phase patterns in ensembles of nonlinear oscillators~\cite{zlotnik2016phase}.
More recently, phase reduction theory has been extended from individual oscillators to networks of coupled dynamical elements exhibiting collective oscillations, allowing the dynamics of an entire network to be described by a single phase variable, called the collective phase~\cite{nakao2018phase}.
Within this framework, the internetwork coupling between a pair of collectively oscillating networks has been optimized for mutual synchronization~\cite{nakao2021sparse}.

Conventional models of network dynamics typically assume pairwise interactions, in which each link connects only two nodes.
However, many real-world systems involve higher-order interactions among three or more units that cannot be decomposed into independent pairwise couplings.
Such interactions arise, for example, in scientific collaborations, ecological communities, brain networks, and social contagion~\cite{battiston2020networks,battiston2021physics}.
Hypergraphs and simplicial complexes have therefore been introduced as natural mathematical frameworks for describing such higher-order networks~\cite{bianconi2021higher,bick2023higher,boccaletti2023structure}.
Higher-order interactions can qualitatively alter collective dynamics~\cite{majhi2022dynamics}, giving rise to phenomena such as multistability~\cite{tanaka2011multistable} and abrupt synchronization transitions~\cite{skardal2020higher,millan2020explosive}, and modifying the stability of synchronized states~\cite{lucas2020multiorder,gambuzza2021stability}.

Studies of synchronization in higher-order networks have shown that both the topology and the directionality of higher-order interactions substantially influence collective dynamics~\cite{lucas2020multiorder,gambuzza2021stability,gallo2022synchronization}.
Accordingly, the control and optimization of synchronization in higher-order networks have recently attracted increasing attention.
For example, Tang \textit{et al.}~\cite{tang2022optimizing} optimized the hypergraph topology of Kuramoto oscillators with higher-order interactions, and Namura \textit{et al.}~\cite{namura2026optimal} used phase reduction theory to design optimal interaction functions that realize higher-order Kuramoto dynamics in systems of limit-cycle oscillators.
These studies, however, have focused on a single network of oscillators, whereas many systems of interest consist of multiple interacting subnetworks whose collective oscillations govern the dynamics of the whole system.
For such collectively oscillating networks, synchronization has been optimized only for pairwise coupling between two networks~\cite{nakao2021sparse}.

\begin{figure}[!t]
	\centering
	\includegraphics[width=0.45\textwidth]{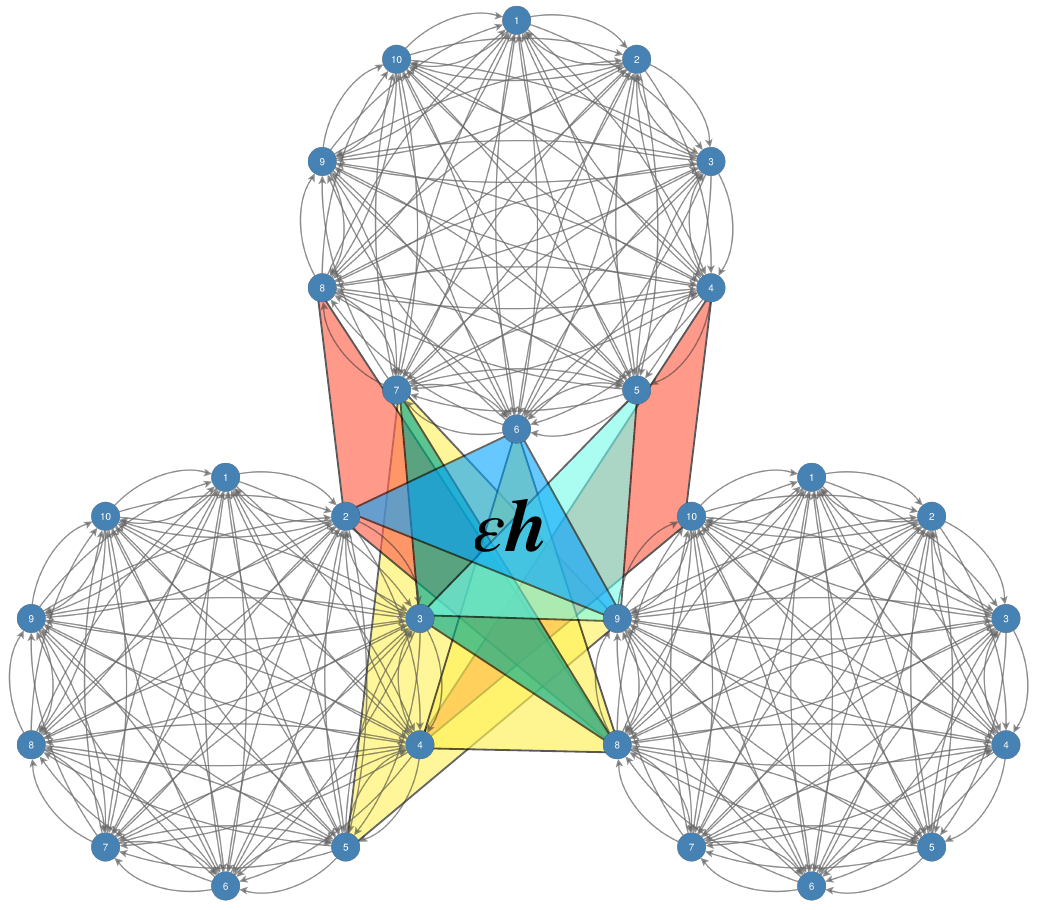}
	\caption{%
		Schematic of three collectively oscillating networks with higher-order internetwork coupling.
		Each network exhibits stable collective oscillations in the absence of internetwork coupling.
		The three networks interact through the internetwork coupling $\epsilon\bm{h}$, which is assumed to be weak compared with the intrinsic dynamics of each network.
		Each colored triangle represents a higher-order internetwork coupling among one node from each of the three networks, and the whole set of such couplings is specified by a coupling tensor.
	}
	\label{fig:three-networks}
\end{figure}

In this study, we consider three collectively oscillating networks interacting through weak higher-order internetwork coupling, as schematically illustrated in figure~\ref{fig:three-networks}, and investigate the optimal design of this coupling.
By extending the phase reduction framework for collective oscillations~\cite{nakao2018phase}, we reduce the full network dynamics to a two-dimensional system of phase-difference equations.
In the reduced system, both the linear stability of the phase-locked state and the rotational property of the phase differences around it are determined by a higher-order coupling tensor shared by the three networks.
Following the optimization approaches developed in Ref.~\cite{nakao2021sparse}, we formulate optimization problems for this coupling tensor for a fixed overall strength of the internetwork coupling and analytically derive the optimal tensors that maximize either the linear stability or the rotational property.
As a numerical example, we consider three collectively oscillating networks, each consisting of four FitzHugh--Nagumo elements coupled in a tree structure.
Numerical simulations of the reduced phase-difference equations show that the two optimized couplings yield, respectively, faster convergence to the phase-locked state and stronger rotational motion of the phase differences around it than a random coupling tensor with the same overall strength.

\section{Model}

In this section, we introduce a general model of three collectively oscillating networks with higher-order internetwork coupling, derive coupled equations for their collective phases by phase reduction and averaging, and obtain a two-dimensional system of phase-difference equations for analyzing the linear stability of the phase-locked state and the rotational property of the phase differences around it.

\subsection{Three collectively oscillating networks}
We first introduce a general model of the three collectively oscillating networks, which can be non-identical in their internal structures and in the dynamics of their individual elements.
We label the networks by $X\in\{A,B,C\}$ and assume that network $X$ consists of $N^X$ dynamical elements.
The state of the $i$th element in network $X$ is denoted by $\bm{x}_i^X\in\R^{m_i^X}$, where $m_i^X$ is the dimension of its state space.
The full dynamical system is given by
\begin{align}
	\dot{\bm{x}}_i^A
	&=
	\bm{f}_i^A(\bm{x}_i^A)
	+\sum_{j=1}^{N^A}
	\bm{g}_{ij}^A(\bm{x}_i^A,\bm{x}_j^A)
	+\epsilon
	\sum_{j=1}^{N^B}\sum_{k=1}^{N^C}
	\bm{h}_{ijk}^A(\bm{x}_i^A,\bm{x}_j^B,\bm{x}_k^C),
	&& i=1,\ldots,N^A,
	\label{eq:model_A}
	\\
	\dot{\bm{x}}_i^B
	&=
	\bm{f}_i^B(\bm{x}_i^B)
	+\sum_{j=1}^{N^B}
	\bm{g}_{ij}^B(\bm{x}_i^B,\bm{x}_j^B)
	+\epsilon
	\sum_{j=1}^{N^C}\sum_{k=1}^{N^A}
	\bm{h}_{ijk}^B(\bm{x}_i^B,\bm{x}_j^C,\bm{x}_k^A),
	&& i=1,\ldots,N^B,
	\label{eq:model_B}
	\\
	\dot{\bm{x}}_i^C
	&=
	\bm{f}_i^C(\bm{x}_i^C)
	+\sum_{j=1}^{N^C}
	\bm{g}_{ij}^C(\bm{x}_i^C,\bm{x}_j^C)
	+\epsilon
	\sum_{j=1}^{N^A}\sum_{k=1}^{N^B}
	\bm{h}_{ijk}^C(\bm{x}_i^C,\bm{x}_j^A,\bm{x}_k^B),
	&& i=1,\ldots,N^C,
	\label{eq:model_C}
\end{align}
where the function $\bm{f}_i^X:\R^{m_i^X}\to\R^{m_i^X}$ describes the intrinsic dynamics of the $i$th element in network $X$, and $\bm{g}_{ij}^X:\R^{m_i^X}\times\R^{m_j^X}\to\R^{m_i^X}$ describes the influence of element $j$ on element $i$ within network $X$.
Since the self-interaction term $\bm{g}_{ii}^X$ can be absorbed into the intrinsic dynamics $\bm{f}_i^X$, we set $\bm{g}_{ii}^X=\bm{0}$ without loss of generality.
For each cyclic permutation $(X,Y,Z)\in\{(A,B,C),(B,C,A),(C,A,B)\}$, the function $\bm{h}_{ijk}^X:\R^{m_i^X}\times\R^{m_j^Y}\times\R^{m_k^Z}\to\R^{m_i^X}$ describes the three-body internetwork coupling through which element $j$ in network $Y$ and element $k$ in network $Z$ jointly influence element $i$ in network $X$.
The small parameter $\epsilon$ ($0<\epsilon\ll1$) characterizes the order of magnitude of the internetwork coupling.
We assume that, in the absence of internetwork coupling, i.e., when $\epsilon=0$, each network possesses an exponentially stable collective limit-cycle solution
$\tilde{\bm{x}}^X(t)=\{\tilde{\bm{x}}_i^X(t)\}_{i=1}^{N^X}$
with period $T^X$ and collective frequency $\omega^X=2\pi/T^X$, satisfying
$\tilde{\bm{x}}_i^X(t+T^X)=\tilde{\bm{x}}_i^X(t)$
for all $t$ and $i=1,\ldots,N^X$.

\subsection{Phase reduction and averaging}

Assuming that the internetwork coupling is weak and that the collective frequencies of the three networks are sufficiently close to each other, we apply collective phase reduction and averaging to derive coupled equations for the collective phases of the three networks.

Following the collective phase reduction framework in Ref.~\cite{nakao2018phase}, we introduce an asymptotic phase function
\begin{equation}
	\Theta^X:
	\mathcal{B}^X
	\to [0,2\pi),
\end{equation}
where $\mathcal{B}^X\subset\R^{m_1^X}\times\cdots\times\R^{m_{N^X}^X}$ denotes the basin of attraction of the collective limit cycle of network $X$.
This function introduces a collective phase
\begin{equation}
	\theta^X
	=
	\Theta^X\left(\{\bm{x}_i^X\}_{i=1}^{N^X}\right)
\end{equation}
for the network state $\{\bm{x}_i^X\}_{i=1}^{N^X}$ such that, in the absence of internetwork coupling, it increases at a constant rate $\omega^X$, i.e., $\dot{\theta}^X=\omega^X$.
We denote the state of network $X$ on its collective limit cycle, parameterized by the collective phase $\theta^X$, as $\{\bar{\bm{x}}_i^X(\theta^X)\}_{i=1}^{N^X}$.
With an appropriate choice of the time origin, this state is related to the collective limit-cycle solution introduced in the previous subsection by $\bar{\bm{x}}_i^X(\theta)=\tilde{\bm{x}}_i^X(\theta/\omega^X)$, and each $\bar{\bm{x}}_i^X(\theta)$ is $2\pi$-periodic in $\theta$.
The collective phase sensitivity function of the $i$th element in network $X$ is defined as the gradient of the asymptotic phase function with respect to $\bm{x}_i^X$, evaluated on the collective limit cycle,
\begin{equation}
	\bm{Q}_i^X(\theta)
	=
	\left.
	\frac{\partial\Theta^X}{\partial\bm{x}_i^X}
	\right|_{\{\bm{x}_j^X\}_{j=1}^{N^X}
		=\{\bar{\bm{x}}_j^X(\theta)\}_{j=1}^{N^X}},
	\qquad
	i=1,\ldots,N^X.
	\label{eq:PSF}
\end{equation}
The collective phase sensitivity functions can be computed numerically by the adjoint method described in Ref.~\cite{nakao2018phase}.

Since the internetwork coupling is weak, we assume that the state of each network remains in the vicinity of its collective limit cycle.
The collective phase of network $X$ then obeys
\begin{equation}
	\dot{\theta}^X
	=
	\omega^X
	+
	\epsilon
	\sum_{i=1}^{N^X}
	\sum_{j=1}^{N^Y}
	\sum_{k=1}^{N^Z}
	\bm{Q}_i^X(\theta^X)\cdot
	\bm{h}_{ijk}^X\!\left(
	\bar{\bm{x}}_i^X(\theta^X),
	\bar{\bm{x}}_j^Y(\theta^Y),
	\bar{\bm{x}}_k^Z(\theta^Z)
	\right)
	+
	O(\epsilon^2)
	\label{eq:phase_reduced}
\end{equation}
for each cyclic permutation $(X,Y,Z)\in\{(A,B,C),(B,C,A),(C,A,B)\}$.

We also assume that the differences among the collective frequencies of the three networks are of order $\epsilon$ and write
\begin{equation}
	\omega^X
	=
	\omega+\epsilon\Delta^X,
	\qquad
	X\in\{A,B,C\},
	\label{eq:detuning}
\end{equation}
where $\omega$ is a reference frequency, chosen, for example, as $\omega=(\omega^A+\omega^B+\omega^C)/3$, and $\epsilon\Delta^X$ represents the frequency detuning of network $X$ from $\omega$.
We then introduce the slow phase
\begin{equation}
	\Phi^X=\theta^X-\omega t,
\end{equation}
which evolves at a rate of $O(\epsilon)$ because the common fast rotation $\omega t$ has been removed.
Since $\Phi^X$ can be regarded as approximately constant over one period of the fast phase $\psi=\omega t$, the $O(\epsilon)$ terms in equation~\eqref{eq:phase_reduced} can be averaged with respect to $\psi$.

We introduce the phase differences
\begin{equation}
	\phi_{XY}=\theta^Y-\theta^X,
	\qquad
	X,Y\in\{A,B,C\}.
	\label{eq:phase_diff}
\end{equation}
For each cyclic permutation $(X,Y,Z)$, the contribution of the three-body internetwork coupling among element $i$ in network $X$, element $j$ in network $Y$, and element $k$ in network $Z$ to the collective phase dynamics of network $X$ is given by the individual phase coupling function
\begin{equation}
	\gamma_{ijk}^X(\phi_{XY},\phi_{XZ})
	=
	\frac{1}{2\pi}
	\int_0^{2\pi}
	\bm{Q}_i^X(\psi)\cdot
	\bm{h}_{ijk}^X\!\left(
	\bar{\bm{x}}_i^X(\psi),
	\bar{\bm{x}}_j^Y(\psi+\phi_{XY}),
	\bar{\bm{x}}_k^Z(\psi+\phi_{XZ})
	\right)
	\,\dd\psi,
	\label{eq:gamma_elem}
\end{equation}
where $i=1,\ldots,N^X$, $j=1,\ldots,N^Y$, and $k=1,\ldots,N^Z$.
Summing these contributions over all combinations of elements yields the effective collective phase coupling function
\begin{equation}
	\Gamma^X(\phi_{XY},\phi_{XZ})
	=
	\sum_{i=1}^{N^X}
	\sum_{j=1}^{N^Y}
	\sum_{k=1}^{N^Z}
	\gamma_{ijk}^X(\phi_{XY},\phi_{XZ}).
	\label{eq:Gamma}
\end{equation}

For example, the averaged equation for the slow phase of network $A$ is
\begin{equation}
	\dot{\Phi}^A
	=
	\epsilon
	\left[
	\Delta^A
	+
	\Gamma^A
	\left(
	\Phi^B-\Phi^A,
	\Phi^C-\Phi^A
	\right)
	\right]
	+
	O(\epsilon^2),
\end{equation}
and the corresponding equations for networks $B$ and $C$ are obtained by cyclic permutation.

Returning to the original collective phases and neglecting the $O(\epsilon^2)$ terms, we obtain the following phase equations,
\begin{align}
	\dot{\theta}^A
	&=
	\omega^A
	+
	\epsilon
	\Gamma^A
	\left(
	\theta^B-\theta^A,
	\theta^C-\theta^A
	\right),
	\label{eq:phase_A}
	\\
	\dot{\theta}^B
	&=
	\omega^B
	+
	\epsilon
	\Gamma^B
	\left(
	\theta^C-\theta^B,
	\theta^A-\theta^B
	\right),
	\label{eq:phase_B}
	\\
	\dot{\theta}^C
	&=
	\omega^C
	+
	\epsilon
	\Gamma^C
	\left(
	\theta^A-\theta^C,
	\theta^B-\theta^C
	\right).
	\label{eq:phase_C}
\end{align}
In general, the collective limit cycles, the collective phase sensitivity functions, and the three-body internetwork coupling functions may differ among the three networks.
Consequently, the effective collective phase coupling functions $\Gamma^A$, $\Gamma^B$, and $\Gamma^C$ need not be identical.

\subsection{Phase-difference equations and linear stability analysis}
\label{sec:phase-diff}

Among these phase differences $\phi_{XY}=\theta^Y-\theta^X$ introduced in equation~\eqref{eq:phase_diff}, only two are independent, and we take them as
\begin{equation}
	\phi_1=\phi_{AB}=\theta^B-\theta^A,
	\qquad
	\phi_2=\phi_{AC}=\theta^C-\theta^A.
	\label{eq:phasediff_def}
\end{equation}
Subtracting equation~\eqref{eq:phase_A} from equations~\eqref{eq:phase_B} and \eqref{eq:phase_C} and using equation~\eqref{eq:detuning} together with $\phi_{BC}=\phi_2-\phi_1$, we obtain the phase-difference equations
\begin{align}
	\dot{\phi}_1&=\epsilon \left[
	\Delta^B-\Delta^A
	+\Gamma^B(\phi_2-\phi_1,-\phi_1)
	-\Gamma^A(\phi_1,\phi_2)\right],\label{eq:phasediff_1}\\
	\dot{\phi}_2&=\epsilon\left[
	\Delta^C-\Delta^A
	+\Gamma^C(-\phi_2,\phi_1-\phi_2)
	-\Gamma^A(\phi_1,\phi_2)\right].\label{eq:phasediff_2}
\end{align}
Thus, the synchronization dynamics of the three collectively oscillating networks is reduced to a two-dimensional system of phase-difference equations.
Let $(\phi_1^*,\phi_2^*)$ be a fixed point of equations~\eqref{eq:phasediff_1} and \eqref{eq:phasediff_2}, corresponding to a phase-locked state of the three networks.
We denote by $\partial_1$ and $\partial_2$ the partial derivatives of a function with respect to its first and second arguments, respectively, and define the derivatives of the effective collective phase coupling functions evaluated at the fixed point as
\begin{align}
	\Gamma_1^A&=\partial_1\Gamma^A(\phi_1^*,\phi_2^*),
	&\Gamma_2^A&=\partial_2\Gamma^A(\phi_1^*,\phi_2^*),\label{eq:derivs_A}\\
	\Gamma_1^B&=\partial_1\Gamma^B(\phi_2^*-\phi_1^*,-\phi_1^*),
	&\Gamma_2^B&=\partial_2\Gamma^B(\phi_2^*-\phi_1^*,-\phi_1^*),\label{eq:derivs_B}\\
	\Gamma_1^C&=\partial_1\Gamma^C(-\phi_2^*,\phi_1^*-\phi_2^*),
	&\Gamma_2^C&=\partial_2\Gamma^C(-\phi_2^*,\phi_1^*-\phi_2^*).\label{eq:derivs_C}
\end{align}
Let $F_1$ and $F_2$ denote the quantities in the square brackets in equations~\eqref{eq:phasediff_1} and \eqref{eq:phasediff_2}, respectively.
The Jacobian matrix $J$ of $(F_1,F_2)$ with respect to $(\phi_1,\phi_2)$, evaluated at the fixed point, is given by
\begin{equation}
	J=
	\begin{pmatrix}
		-\Gamma_1^B-\Gamma_2^B-\Gamma_1^A & \Gamma_1^B-\Gamma_2^A\\
		\Gamma_2^C-\Gamma_1^A & -\Gamma_1^C-\Gamma_2^C-\Gamma_2^A
	\end{pmatrix}.
	\label{eq:J_general}
\end{equation}
Here, the small parameter $\epsilon$ is excluded from the definition of $J$. The Jacobian matrix of the original phase-difference equations~\eqref{eq:phasediff_1} and \eqref{eq:phasediff_2} is $\epsilon J$, and the eigenvalues and related quantities introduced below are defined with respect to the rescaled time $\epsilon t$.
Defining
\begin{equation}
	\begin{aligned}
		\sigma&=-\Tr J
		=\Gamma_1^A+\Gamma_2^A+\Gamma_1^B+\Gamma_2^B+\Gamma_1^C+\Gamma_2^C,\\
		\tau&=\det J
		=(\Gamma_1^B+\Gamma_2^B+\Gamma_1^A)(\Gamma_1^C+\Gamma_2^C+\Gamma_2^A)
		-(\Gamma_1^B-\Gamma_2^A)(\Gamma_2^C-\Gamma_1^A),
	\end{aligned}
	\label{eq:sigma_tau}
\end{equation}
we obtain the eigenvalues of $J$ as
\begin{equation}
	\lambda_{\pm}=-\frac{\sigma}{2}\pm\frac{\sqrt{\sigma^2-4\tau}}{2}.
	\label{eq:eigen}
\end{equation}

We introduce the linear decay rate and the rotational component as
\begin{equation}
	\Lambda=\frac{\sigma}{2},
	\qquad
	R=\frac{1}{2}\sqrt{\max\{4\tau-\sigma^2,0\}},
	\label{eq:Lambda_R}
\end{equation}
respectively.
In this study, we restrict our attention to the case $4\tau\geq\sigma^2$, in which the two eigenvalues share the common real part $-\Lambda$ and are written as $\lambda_{\pm}=-\Lambda\pm iR$.
These eigenvalues form a complex-conjugate pair for $R>0$ and reduce to a double real eigenvalue $-\Lambda$ for $R=0$.
In this case, $\Lambda$ gives the exponential decay rate of small perturbations from the fixed point, $R$ gives the angular frequency of the rotation of the phase differences around the fixed point, and the fixed point is linearly stable if and only if $\Lambda>0$.
On the original time scale of equations~\eqref{eq:phasediff_1} and \eqref{eq:phasediff_2}, the eigenvalues are $\epsilon\lambda_{\pm}$, so that the decay rate and the rotational angular frequency are $\epsilon\Lambda$ and $\epsilon R$, respectively.

When $4\tau<\sigma^2$, on the other hand, the eigenvalues are real and distinct, and the decay rate of the slower mode, $(\sigma-\sqrt{\sigma^2-4\tau})/2$, is smaller than $\Lambda$, so that $\Lambda$ no longer characterizes the convergence to the fixed point.
We do not consider this case in the following. The optimization beyond this regime is discussed in Appendix~\ref{app:kkt}.

\section{Optimal design of higher-order couplings}
\label{sec:optimization}
In this section, we optimize the strengths of the higher-order (three-body) internetwork couplings among the three networks on the basis of the quantities obtained in the linear stability analysis in the previous section.
We express each three-body coupling as the product of a scalar coefficient and a prescribed coupling function, and label the elements of the coupling tensor $\Ct$ by the index triplet $(i,j,k)$, where the indices $i=1,\ldots,N^A$, $j=1,\ldots,N^B$, and $k=1,\ldots,N^C$ refer to the elements of networks $A$, $B$, and $C$, respectively.
That is, we write the three-body coupling functions as
\begin{equation}
	\begin{aligned}
		\bm{h}_{ijk}^A&=\Ct_{ijk}\bm{H}_{ijk}^A,
		&\qquad
		\gamma_{ijk}^A(\varphi,\varphi')&=\Ct_{ijk}V_{ijk}^A(\varphi,\varphi'),\\
		\bm{h}_{jki}^B&=\Ct_{ijk}\bm{H}_{jki}^B,
		&\qquad
		\gamma_{jki}^B(\varphi,\varphi')&=\Ct_{ijk}V_{jki}^B(\varphi,\varphi'),\\
		\bm{h}_{kij}^C&=\Ct_{ijk}\bm{H}_{kij}^C,
		&\qquad
		\gamma_{kij}^C(\varphi,\varphi')&=\Ct_{ijk}V_{kij}^C(\varphi,\varphi'),
	\end{aligned}
	\label{eq:decomposition}
\end{equation}
where $\bm{H}_{ijk}^A$, $\bm{H}_{jki}^B$, and $\bm{H}_{kij}^C$ are the prescribed coupling functions, and the individual phase coupling functions are decomposed accordingly, with $V_{ijk}^A$, $V_{jki}^B$, and $V_{kij}^C$ obtained by replacing $\bm{h}_{ijk}^A$, $\bm{h}_{jki}^B$, and $\bm{h}_{kij}^C$ in equation~\eqref{eq:gamma_elem} with these prescribed functions.
The same coefficient $\Ct_{ijk}$ thus weights the coupling functions with the local index orders $(i,j,k)$, $(j,k,i)$, and $(k,i,j)$ in networks $A$, $B$, and $C$, respectively; that is, the three networks share a single coupling tensor $\Ct$ of size $N^A\times N^B\times N^C$.
Consequently, for fixed phase differences, $\Gamma^X$ is a linear function of $\Ct$.

In what follows, for notational simplicity, we introduce a vector representation of the coupling tensor.
We arrange the elements $\Ct_{ijk}$ in a fixed order of $(i,j,k)$ and define the coupling vector $\bm{c}=\mathrm{vec}(\Ct)\in\R^L$ with $L=N^AN^BN^C$.
Denoting by $\ell(i,j,k)$ the position of $\Ct_{ijk}$ in $\bm{c}$, we define the gradient vectors consisting of the partial derivatives of $V_{ijk}^A$, $V_{jki}^B$, and $V_{kij}^C$ in equation~\eqref{eq:decomposition} as
\begin{equation}
	\begin{aligned}
		(\bm{a}_1^{A})_{\ell(i,j,k)}&=\partial_1V_{ijk}^A(\phi_1^*,\phi_2^*),
		&\qquad
		(\bm{a}_2^{A})_{\ell(i,j,k)}&=\partial_2V_{ijk}^A(\phi_1^*,\phi_2^*),\\
		(\bm{a}_1^{B})_{\ell(i,j,k)}&=\partial_1V_{jki}^B(\phi_2^*-\phi_1^*,-\phi_1^*),
		&\qquad
		(\bm{a}_2^{B})_{\ell(i,j,k)}&=\partial_2V_{jki}^B(\phi_2^*-\phi_1^*,-\phi_1^*),\\
		(\bm{a}_1^{C})_{\ell(i,j,k)}&=\partial_1V_{kij}^C(-\phi_2^*,\phi_1^*-\phi_2^*),
		&\qquad
		(\bm{a}_2^{C})_{\ell(i,j,k)}&=\partial_2V_{kij}^C(-\phi_2^*,\phi_1^*-\phi_2^*).
	\end{aligned}
	\label{eq:grad_vectors}
\end{equation}
With these definitions, the derivatives $\Gamma_m^X$ defined in equations~\eqref{eq:derivs_A}--\eqref{eq:derivs_C} are expressed as
\begin{equation}
	\Gamma_m^X=(\bm{a}_m^X)^T\bm{c},
	\qquad
	m\in\{1,2\},
	\quad
	X\in\{A,B,C\},
	\label{eq:Gamma_linear}
\end{equation}
i.e., $\bm{a}_m^X$ is the gradient of $\Gamma_m^X$ with respect to the coupling vector $\bm{c}$.

Defining
\begin{equation}
	\bm{w}=\sum_{X\in\{A,B,C\}}\left(\bm{a}_1^{X}+\bm{a}_2^{X}\right),
	\label{eq:w}
\end{equation}
we can express $\sigma$ in equation~\eqref{eq:sigma_tau} as the linear form
\begin{equation}
	\sigma=\bm{w}^T\bm{c}.
	\label{eq:sigma_linear}
\end{equation}
Similarly, with $\mathrm{sym}(K)=(K+K^T)/2$ for a square matrix $K$, we define the symmetric matrix
\begin{align}
	M=\mathrm{sym}\Big(
	&\bm{a}_1^{A}(\bm{a}_1^{B})^T
	+\bm{a}_1^{B}(\bm{a}_1^{C})^T
	+\bm{a}_1^{C}(\bm{a}_1^{A})^T\notag\\
	&+\bm{a}_2^{A}(\bm{a}_2^{B})^T
	+\bm{a}_2^{B}(\bm{a}_2^{C})^T
	+\bm{a}_2^{C}(\bm{a}_2^{A})^T\notag\\
	&+\bm{a}_1^{C}(\bm{a}_2^{B})^T
	+\bm{a}_1^{B}(\bm{a}_2^{A})^T
	+\bm{a}_1^{A}(\bm{a}_2^{C})^T
	\Big),
	\label{eq:M}
\end{align}
so that the determinant $\tau$ in equation~\eqref{eq:sigma_tau} is expressed as the quadratic form
\begin{equation}
	\tau=\bm{c}^TM\bm{c}.
	\label{eq:tau_quad}
\end{equation}
Substituting equations~\eqref{eq:sigma_linear} and \eqref{eq:tau_quad} into equation~\eqref{eq:Lambda_R}, we express the linear decay rate and the rotational component in terms of the coupling vector as
\begin{equation}
	\Lambda=\frac{1}{2}\bm{w}^T\bm{c},
	\qquad
	R=\frac{1}{2}\sqrt{\max\{\bm{c}^TP\bm{c},0\}},
	\qquad
	P=4M-\bm{w}\bm{w}^T.
	\label{eq:Lambda_R_c}
\end{equation}
Thus, $\Lambda$ is a linear function and $R^2$ is a quadratic function of $\bm{c}$ in the regime $\bm{c}^TP\bm{c}\geq0$, which corresponds to the condition $4\tau\geq\sigma^2$ considered in this study.
To fix the overall magnitude of the coupling tensor, we impose the norm constraint
\begin{equation}
	\norm{\bm{c}}_2
	= \kappa,
	\label{eq:norm_constraint}
\end{equation}
where $\kappa>0$ is a prescribed constant and
$\norm{\bm{c}}_2
=
\left(\sum_{i=1}^{N^A}\sum_{j=1}^{N^B}\sum_{k=1}^{N^C} (\Ct_{ijk})^2\right)^{1/2}$
is the Euclidean norm of $\bm{c}$, which coincides with the Frobenius norm $\norm{\Ct}_F$ of the tensor $\Ct$.

We formulate optimization problems for the coupling vector $\bm{c}$, or equivalently the tensor $\Ct$, that maximize either the linear decay rate $\Lambda$ or the rotational component $R$ in equation~\eqref{eq:Lambda_R_c}, evaluated at the fixed point $(\phi_1^*,\phi_2^*)$, subject to the norm constraint~\eqref{eq:norm_constraint}.

\subsection{Optimization of linear stability}
\label{sec:opt_Lambda}

We first optimize the linear stability of the fixed point $(\phi_1^*,\phi_2^*)$.
The problem of maximizing the linear decay rate $\Lambda$ in equation~\eqref{eq:Lambda_R_c} under the norm constraint~\eqref{eq:norm_constraint} is formulated as
\begin{equation}
	\begin{aligned}
		\max_{\bm{c}}\quad & \Lambda(\bm{c})=\frac{1}{2}\bm{w}^{T}\bm{c} \\
		\text{subject to}\quad & \norm{\bm{c}}_2=\kappa.
	\end{aligned}
	\label{eq:opt_Lambda}
\end{equation}
By the Cauchy--Schwarz inequality, $\bm{w}^{T}\bm{c}\leq\norm{\bm{w}}_2\norm{\bm{c}}_2$, with equality if and only if $\bm{c}$ is a nonnegative scalar multiple of $\bm{w}$.
Therefore, assuming $\bm{w}\neq\bm{0}$, the optimal coupling vector is given by
\begin{equation}
	\bm{c}_{\Lambda}^{\star}
	=
	\kappa\frac{\bm{w}}{\norm{\bm{w}}_2},
	\label{eq:linear_fixed_norm}
\end{equation}
and the maximum linear decay rate is
\begin{equation}
	\Lambda^{\max}(\phi_1^*,\phi_2^*)
	=
	\frac{\kappa}{2}\norm{\bm{w}}_2.
	\label{eq:linear_fixed_norm_lst}
\end{equation}
This optimal coupling maximizes the local convergence speed toward the phase-locked state.
Note that $\bm{w}$ depends on the fixed point through equation~\eqref{eq:grad_vectors}, and that this optimization does not by itself guarantee the condition $4\tau\geq\sigma^2$ at $\bm{c}_{\Lambda}^{\star}$, which should therefore be verified for the obtained coupling.
As discussed in Appendix~\ref{app:kkt}, this condition is satisfied if and only if the scalar $s=\bm{w}^TP\bm{w}$ is nonnegative, and the case $s<0$ requires a more general optimization.

\subsection{Optimization of the rotational component}
\label{sec:opt_R}

Next, we consider the maximization of the rotational component $R$.
Under the condition $4\tau\geq\sigma^2$ assumed in section~\ref{sec:phase-diff}, i.e., $\bm{c}^TP\bm{c}\geq0$, we have $R=\sqrt{\bm{c}^TP\bm{c}}/2$, so that maximizing $R$ is equivalent to maximizing the quadratic form $\bm{c}^TP\bm{c}$.
The optimization problem under the norm constraint~\eqref{eq:norm_constraint} is thus formulated as
\begin{equation}
	\begin{aligned}
		\max_{\bm{c}}\quad & \bm{c}^TP\bm{c} \\
		\text{subject to}\quad & \norm{\bm{c}}_2=\kappa.
	\end{aligned}
	\label{eq:opt_R}
\end{equation}
Let $\nu_{\max}$ be the largest eigenvalue of the symmetric matrix $P$ and $\bm{v}_{\max}$ a corresponding unit eigenvector.
Since problem~\eqref{eq:opt_R} is equivalent to the maximization of a Rayleigh quotient, its maximum is $\kappa^2\nu_{\max}$, attained at $\bm{c}=\pm\kappa\bm{v}_{\max}$.
Hence, when $\nu_{\max}>0$, the optimal coupling vector is
\begin{equation}
	\bm{c}_R^{\star}=\kappa\bm{v}_{\max},
	\label{eq:rot_fixed_norm}
\end{equation}
which attains the maximum rotational component
\begin{equation}
	R^{\max}(\phi_1^*,\phi_2^*)=\frac{\kappa}{2}\sqrt{\nu_{\max}},
	\label{eq:rot_fixed_norm_rot}
\end{equation}
i.e., the maximum angular frequency of the rotation of the phase differences around the fixed point.
The two signs give the same $R$, since $\bm{c}^TP\bm{c}$ is invariant under $\bm{c}\to-\bm{c}$, whereas $\Lambda(\bm{c})$ changes sign.
We therefore choose the sign of $\bm{v}_{\max}$ such that $\bm{w}^{T}\bm{v}_{\max}\geq0$, which gives $\Lambda(\bm{c}_R^{\star})=\kappa\bm{w}^{T}\bm{v}_{\max}/2\geq0$.
Because $\bm{c}^TP\bm{c}=\kappa^2\nu_{\max}>0$ at $\bm{c}_R^{\star}$, the optimal coupling automatically satisfies $4\tau>\sigma^2$, i.e., the eigenvalues of $J$ form a complex-conjugate pair.
Note that this optimization guarantees neither the linear stability of the fixed point, which holds if and only if $\bm{w}^{T}\bm{v}_{\max}>0$, nor the condition $\nu_{\max}>0$ itself, without which $R=0$ for every coupling vector and no rotational motion can be generated.
Both cases are discussed in Appendix~\ref{app:rotation}.

\subsection{Random coupling}
\label{sec:random}
To evaluate the optimization results, we also consider, as a baseline, random coupling vectors with the same overall strength as the optimal ones.
To satisfy the norm constraint~\eqref{eq:norm_constraint}, we sample each element of $\bm{c}$ independently from the standard normal distribution and normalize the resulting vector,
\begin{equation}
	\bm{c}_{\mathrm{rand}}
	=
	\kappa\frac{\bm{r}}{\norm{\bm{r}}_2},
	\qquad
	r_{\ell}\sim\mathcal{N}(0,1),
	\quad
	\ell=1,\ldots,L.
	\label{eq:random}
\end{equation}
In section~\ref{sec:result_R}, we numerically obtain a fixed point of equations~\eqref{eq:phasediff_1} and \eqref{eq:phasediff_2} for each sampled coupling vector and retain only those for which the fixed point is a stable spiral, i.e., $\Lambda>0$ and $4\tau>\sigma^2$.

In section~\ref{sec:result_Lambda}, to compare the convergence to the in-phase phase-locked state $(0,0)$, we use a random coupling tensor invariant under cyclic permutations of its indices, $\Ct_{ijk}=\Ct_{jki}=\Ct_{kij}$, for which $(0,0)$ is a fixed point (see Appendix~\ref{sec:appendix_identical}).
For this purpose, we replace the random vector $\bm{r}$ in equation~\eqref{eq:random} with its average over cyclic permutations of the indices,
\begin{equation}
	\bar{r}_{\ell(i,j,k)}=\frac{1}{3}\left(r_{\ell(i,j,k)}+r_{\ell(j,k,i)}+r_{\ell(k,i,j)}\right),
	\label{eq:random_cyclic}
\end{equation}
so that the resulting coupling vector $\bm{c}_{\mathrm{rand}}=\kappa\bar{\bm{r}}/\norm{\bar{\bm{r}}}_2$ is invariant under cyclic permutations.
Since $R=0$ for all such coupling vectors in the networks of section~\ref{sec:result_Lambda}, as shown in Appendix~\ref{app:identical_symmetry}, the fixed point $(0,0)$ is linearly stable if and only if $\Lambda>0$.
Because $\Lambda=\bm{w}^T\bm{c}/2$ changes sign under $\bm{c}\to-\bm{c}$ while $(0,0)$ remains a fixed point, we reverse the sign of $\bm{c}_{\mathrm{rand}}$ whenever $\Lambda<0$.

\subsection{Numerical calculations}
\label{sec:num_calc}

In the optimization problems formulated above, the fixed point
$(\phi_1^*,\phi_2^*)$ has been regarded as given.
However, the fixed point is determined by the effective collective phase
coupling functions $\Gamma^X$, which depend on the coupling vector
$\bm{c}$.
In practice, we therefore compute the optimal coupling vector and the
corresponding fixed point by an alternating iteration as follows.
Given a fixed point $(\phi_1^{(r)},\phi_2^{(r)})$ at the $r$th
iteration ($r=0,1,2,\ldots$), we evaluate the gradient vectors in
equation~\eqref{eq:grad_vectors}, and hence $\bm{w}$ or $P$ in equations~\eqref{eq:w} and \eqref{eq:Lambda_R_c}, at this point, and compute the coupling vector $\bm{c}^{(r)}$ from
equation~\eqref{eq:linear_fixed_norm} or \eqref{eq:rot_fixed_norm},
respectively.
We then construct the coupling functions $\Gamma^X$ from $\bm{c}^{(r)}$ and apply a root-finding method to equations~\eqref{eq:phasediff_1} and \eqref{eq:phasediff_2},
using $(\phi_1^{(r)},\phi_2^{(r)})$ as the initial guess
to obtain the next fixed point $(\phi_1^{(r+1)},\phi_2^{(r+1)})$.
Starting from a given initial point $(\phi_1^{(0)},\phi_2^{(0)})$, we repeat this alternating update until the coupling vector and the fixed point become self-consistent, that is, until
$\sqrt{(\Delta\phi_1^{(r)})^2+(\Delta\phi_2^{(r)})^2}<\eta$
with $\eta=10^{-8}$, where
$\Delta\phi_i^{(r)}=\phi_i^{(r+1)}-\phi_i^{(r)}$ is taken modulo $2\pi$
in $[-\pi,\pi)$.
When this criterion is satisfied at $r=r^{*}$, we adopt
$(\phi_1^{(r^{*}+1)},\phi_2^{(r^{*}+1)})$ as the fixed point and
the coupling vector $\bm{c}^{(r^{*}+1)}$ evaluated at this fixed point
as the optimized coupling vector.
Finally, we compute the eigenvalues of the Jacobian at the obtained
fixed point and confirm that it is linearly stable.
In the optimization of the rotational component, we further confirm
that it is a stable spiral, i.e., that the eigenvalues are complex
conjugates with negative real parts.

We note that this iteration is not necessary when the three networks
are identical and their three-body coupling functions are also common,
as in section~\ref{sec:result_Lambda}.
In this case, the three effective collective phase coupling functions
coincide, $\Gamma^A=\Gamma^B=\Gamma^C$, and the right-hand sides of
equations~\eqref{eq:phasediff_1} and \eqref{eq:phasediff_2} vanish at
$(\phi_1,\phi_2)=(0,0)$, as shown in Appendix~\ref{sec:appendix_identical};
that is, the in-phase phase-locked state is a fixed point.
Therefore, starting from $(0,0)$, the fixed point stays at $(0,0)$ and
the gradient vectors need not be updated.

For non-identical networks in section~\ref{sec:result_R}, there is no
canonical choice of the origins of the three collective phases
$\theta^X$.
Since $\phi_1$ and $\phi_2$ are differences of the collective phases,
shifting the phase origins merely translates the fixed point
$(\phi_1^*,\phi_2^*)$ without changing the Jacobian matrix $J$, and
hence $\Lambda$ and $R$, so the exact location of the fixed point has
no special meaning.
Therefore, the values of $\Lambda$ and $R$ obtained for different
coupling vectors with different fixed points can be directly compared,
as shown later in table~\ref{tab:results} and
figure~\ref{fig:opt_R}.
However, the iteration may converge to different self-consistent pairs
of the coupling vector and the fixed point, yielding different values
of $\Lambda$ or $R$, depending on the initial point.
In the optimization of the rotational component, we therefore perform the iteration from multiple initial points.
Specifically, we use $(0,0)$ and up to five additional initial points.
The additional points are chosen from a $64\times64$ grid on the
phase-difference plane as those at which the largest eigenvalue
$\nu_{\max}$ of $P$ takes large positive values, under the
condition that they are separated from each other by at least $\pi/4$.
Among the converged results whose fixed points are stable spirals, we adopt the one giving the largest $R$.
Note that this procedure does not guarantee the global optimum, and
selecting the most suitable solution in a systematic way is beyond the
scope of this study.

\section{Example: collectively oscillating tree networks}
\label{sec:example}

As a numerical example, we consider three collectively oscillating networks, each consisting of four FitzHugh--Nagumo elements coupled in a tree structure.
The state of element $i$ in network $X$ is represented by $\bm{x}_i^X=(u_i^X,v_i^X)^T\in\R^2$, i.e., $m_i^X=2$ for all $i$, and the dynamics of each element is described by
\begin{equation}
	\bm{f}_i^X(\bm{x}_i^X)
	=
	\begin{pmatrix}
		\delta\left(a+v_i^X-bu_i^X\right)\\
		v_i^X-(v_i^X)^3/3-u_i^X+I_i
	\end{pmatrix},
	\label{eq:FHN}
\end{equation}
where $\delta$, $a$, $b$, and $I_i$ are parameters.
We fix $\delta=0.08$, $a=0.7$, and $b=0.8$ in the following calculations.
Depending on $I_i$, each element exhibits oscillatory or excitable dynamics; here, the root element is oscillatory with $I_1=0.8$ and the three leaf elements are excitable with $I_2=I_3=I_4=0.2$.
The intranetwork coupling is assumed to be diffusive through the $v$ components,
\begin{equation}
	\bm{g}_{ij}^X(\bm{x}_i^X,\bm{x}_j^X)
	=
	K_{ij}^X
	\begin{pmatrix}
		0\\
		v_j^X-v_i^X
	\end{pmatrix},
	\label{eq:intra}
\end{equation}
where $K_{ij}^X$ is the $(i,j)$ component of the coupling matrix $K^X=\{K_{ij}^X\}$ representing the weights of the diffusive coupling within network $X$.
We use the tree coupling matrix
\begin{equation}
	K=\begin{pmatrix}
		0&0.5&0.5&0.5\\
		1&0&0&0\\
		1&0&0&0\\
		1&0&0&0
	\end{pmatrix},
	\label{eq:K}
\end{equation}
in which the coupling from the leaf elements to the root element is half as strong as that in the opposite direction.
We consider two settings for the three networks.
For the optimization of the linear stability, we use identical networks, $K^A=K^B=K^C=K$.
For the optimization of the rotational component, we use non-identical networks, $K^A=K$, $K^B=0.9K$, and $K^C=1.1K$.
In both settings, the three networks share a single higher-order coupling tensor $\Ct$ as assumed in section~\ref{sec:optimization}.
For the three-body internetwork coupling, we use the basis
\begin{equation}
	\bm{H}_{ijk}^X(\bm{x}_i^X,\bm{x}_j^Y,\bm{x}_k^Z)
	=
	\begin{pmatrix}
		0\\
		\sin\left(v_j^Y+v_k^Z-2v_i^X\right)
	\end{pmatrix}
	\label{eq:H_basis}
\end{equation}
for each cyclic permutation $(X,Y,Z)$.
This coupling vanishes when the three elements have equal $v$ components.
The overall scale of the internetwork coupling is $\epsilon=0.01$, and the coupling tensors are normalized as $\norm{\Ct}_F=\kappa=3.0$.

\begin{figure}
	\centering
	\includegraphics[width=0.7\textwidth]{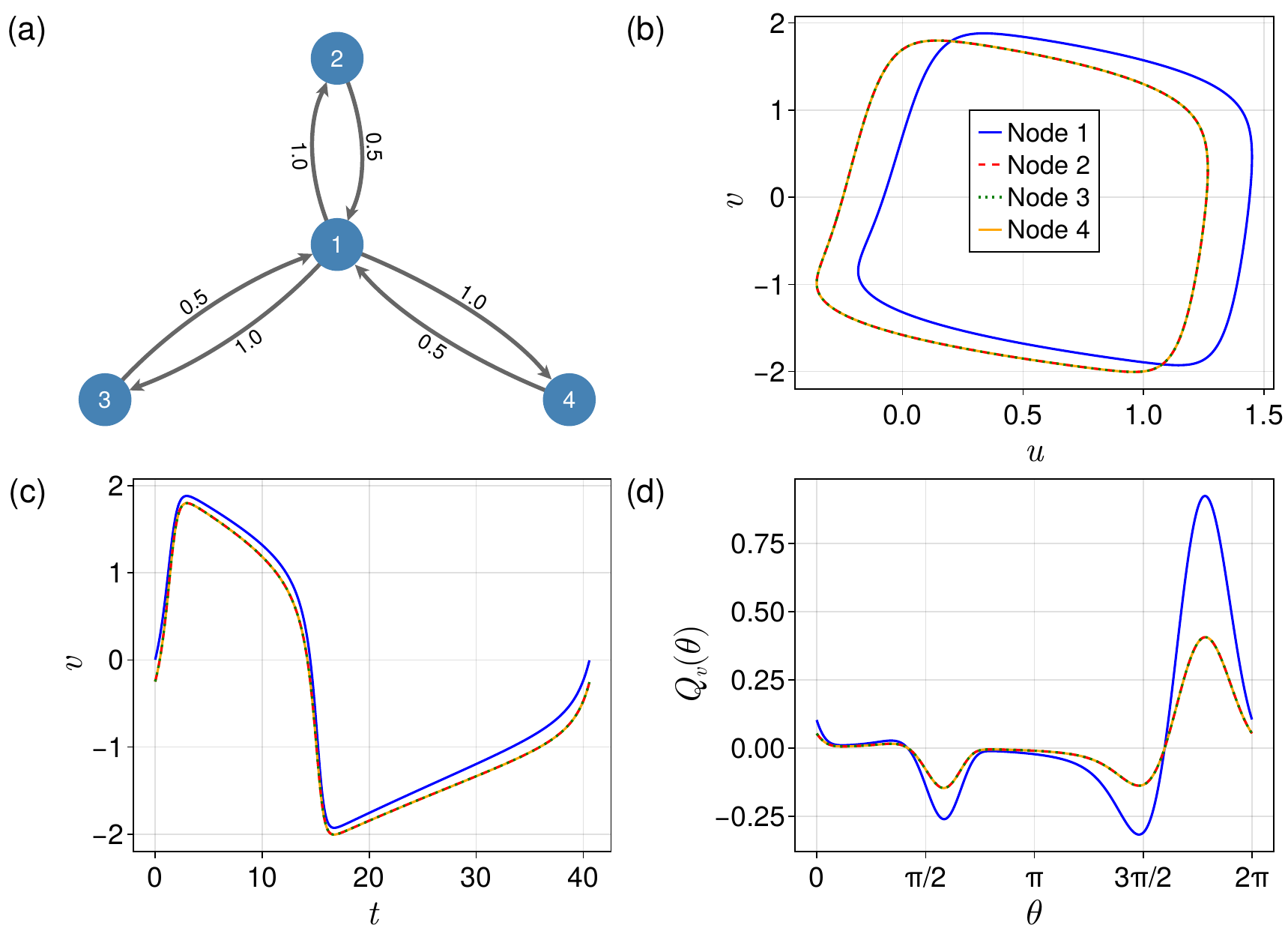}
	\caption{%
		Collective oscillation of a tree network of four FitzHugh--Nagumo elements.
		(a) Network topology.
		(b) Collective limit cycle in the $(u,v)$ plane.
		(c) Time evolution of the $v$ components over one period.
		(d) The $v$ components of the collective phase sensitivity functions.
		The three leaf elements are equivalent, so their curves coincide in (b)--(d).
	}
	\label{fig:tree-network}
\end{figure}

Figure~\ref{fig:tree-network} shows the tree network and its collective oscillation.
Figure~\ref{fig:tree-network}(a) shows the network topology, in which the root element is connected to the three leaf elements by the intranetwork coupling~\eqref{eq:intra}.
Figure~\ref{fig:tree-network}(b) shows the collective limit-cycle trajectory of the four elements, and figure~\ref{fig:tree-network}(c) shows the time evolution of their $v$ components over one oscillation period.
Figure~\ref{fig:tree-network}(d) shows the $v$ components of the collective phase sensitivity functions $\bm{Q}_i(\theta)$ defined in equation~\eqref{eq:PSF}, obtained numerically by the adjoint method.
The three collectively oscillating networks are coupled through the higher-order internetwork coupling as illustrated in figure~\ref{fig:three-networks}.

\subsection{Optimization of linear stability}
\label{sec:result_Lambda}
\begin{figure} [!t]
	\centering
	\includegraphics[width=0.65\textwidth]{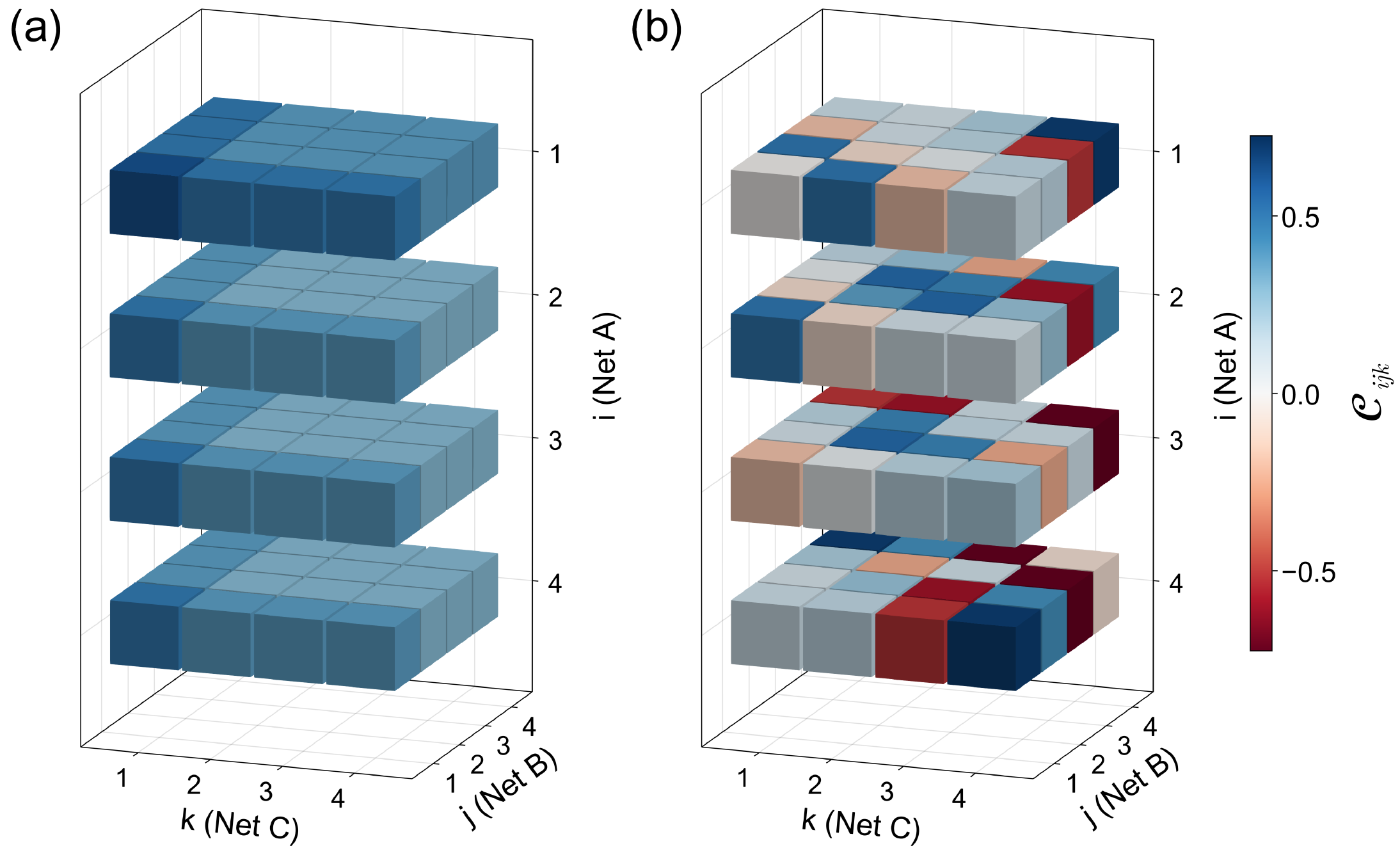}
	\caption{%
		Coupling tensors for the identical networks $K^A=K^B=K^C=K$.
		(a) Optimal coupling tensor $\Ct_\Lambda^\star$ maximizing the linear decay rate and (b) random cyclically invariant coupling tensor $\Ct_{\Lambda,\mathrm{rand}}$ with the same Frobenius norm $\norm{\Ct}_F=\kappa=3.0$.
		Each cube represents the element $\Ct_{ijk}$, where the indices $i$, $j$, and $k$ label the elements of networks $A$, $B$, and $C$, respectively, and its color represents the value of $\Ct_{ijk}$ on the common scale shown on the right.
		Both tensors are invariant under cyclic permutations of the indices, $\Ct_{ijk}=\Ct_{jki}=\Ct_{kij}$.
	}
	\label{fig:tensor_Lambda}
\end{figure}

We first consider the optimization of the linear stability, using the
identical networks $K^A=K^B=K^C=K$ with a coupling tensor invariant
under cyclic permutations of its indices,
$\Ct_{ijk}=\Ct_{jki}=\Ct_{kij}$, so that the three-body coupling
functions are also common to the three networks.
In this case, as shown in Appendix~\ref{sec:appendix_identical}, the
in-phase phase-locked state $(\phi_1^*,\phi_2^*)=(0,0)$ is a fixed
point, and $4\tau\geq\sigma^2$ holds automatically.
Evaluating the general solution~\eqref{eq:linear_fixed_norm} at
$(0,0)$, we obtain the optimal coupling vector
$\bm{c}_\Lambda^\star\propto\bm{w}$.
Since $\bm{w}$ evaluated at $(0,0)$ is invariant under cyclic
permutations of $(i,j,k)$, so is the resulting optimal coupling tensor
$\Ct_\Lambda^\star$.
Hence, $(0,0)$ remains a fixed point for $\Ct_\Lambda^\star$, and the
optimization is consistent with the setting considered here.
The optimal coupling tensor $\Ct_\Lambda^\star$ is shown in
figure~\ref{fig:tensor_Lambda}(a) and explicitly listed in
Appendix~\ref{app:tensors}, and gives $\Lambda=4.6220$ and
$R=0$ at $(0,0)$.
For this coupling, the scalar $s=\bm{w}^TP\bm{w}$ is zero and hence satisfies the condition
$s\geq0$, as shown in Appendix~\ref{app:identical_symmetry}, so that
the analytical solution~\eqref{eq:linear_fixed_norm} is optimal for the actual convergence rate at this fixed point as well.
For comparison, we generate a random coupling tensor
$\Ct_{\Lambda,\mathrm{rand}}$ invariant under cyclic permutations as
described in section~\ref{sec:random}.
This random coupling tensor, shown in figure~\ref{fig:tensor_Lambda}(b)
and explicitly listed in Appendix~\ref{app:tensors}, gives
$\Lambda=1.0342$ and $R=0$ at $(0,0)$, so that the
convergence rates to the same in-phase phase-locked state can be compared directly.

\begin{figure} [!t]
	\centering
	\includegraphics[width=\textwidth]{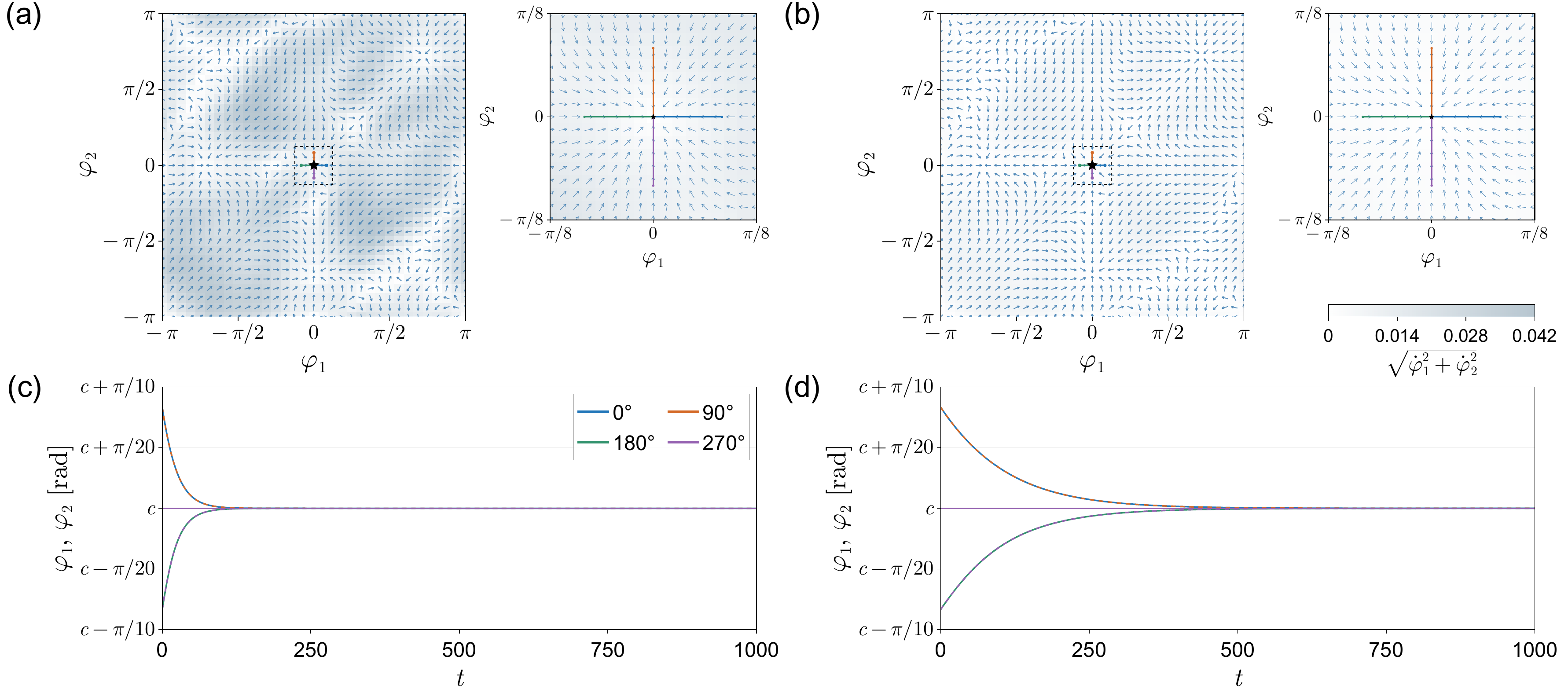}
	\caption{%
		Phase-difference dynamics for the identical networks.
		(a), (b) Vector fields of the phase-difference equations~\eqref{eq:phasediff_1} and \eqref{eq:phasediff_2} for the optimal coupling tensor $\Ct_\Lambda^\star$ and for the random coupling tensor $\Ct_{\Lambda,\mathrm{rand}}$ in figure~\ref{fig:tensor_Lambda}, respectively.
		The arrows show the direction of the flow, the background color shows its speed $\sqrt{\dot{\phi}_1^2+\dot{\phi}_2^2}$ on a common scale, and the star indicates the stable fixed point at the origin $(0,0)$, i.e., the in-phase phase-locked state.
		The right panels of (a) and (b) magnify the dashed square around the origin, where the colored curves show the trajectories starting from $(\pi/12,0)$, $(0,\pi/12)$, $(-\pi/12,0)$, and $(0,-\pi/12)$, which are labeled $0^\circ$, $90^\circ$, $180^\circ$, and $270^\circ$ in the legend in (c), respectively.
		The values of $\Lambda$ and $R$ are listed in table~\ref{tab:results}.
		(c), (d) Time evolution of the phase differences $\phi_1$ (solid lines) and $\phi_2$ (dashed lines) for the trajectories shown in the right panels of (a) and (b), respectively, where the vertical axes are labeled relative to $c=(\phi_1^*+\phi_2^*)/2=0$.
	}
	\label{fig:opt_Lambda}
\end{figure}

Figures~\ref{fig:opt_Lambda}(a) and \ref{fig:opt_Lambda}(b) show the vector fields of the phase-difference equations~\eqref{eq:phasediff_1} and \eqref{eq:phasediff_2} for the two couplings, together with trajectories starting from the four initial phase differences $(\phi_1(0),\phi_2(0))=(\pm\pi/12,0)$ and $(0,\pm\pi/12)$, where the background color represents the speed $\sqrt{\dot{\phi}_1^2+\dot{\phi}_2^2}$ of the phase differences on a common scale, and the right panels magnify the region near the fixed point $(0,0)$.
Figures~\ref{fig:opt_Lambda}(c) and \ref{fig:opt_Lambda}(d) show the time evolution of the phase differences along these trajectories.
The optimal coupling gives a linear decay rate about $4.47$
times larger than that of the random coupling, and the trajectories
correspondingly exhibit faster convergence to the in-phase
phase-locked state.
For both couplings, the rotational component vanishes, since $R=0$ for every cyclically invariant coupling tensor in the present tree networks, as shown in Appendix~\ref{app:identical_symmetry}, and the approach to the in-phase phase-locked state is therefore non-oscillatory.

\subsection{Optimization of the rotational component}
\label{sec:result_R}
\begin{figure} [!t]
	\centering
	\includegraphics[width=0.65\textwidth]{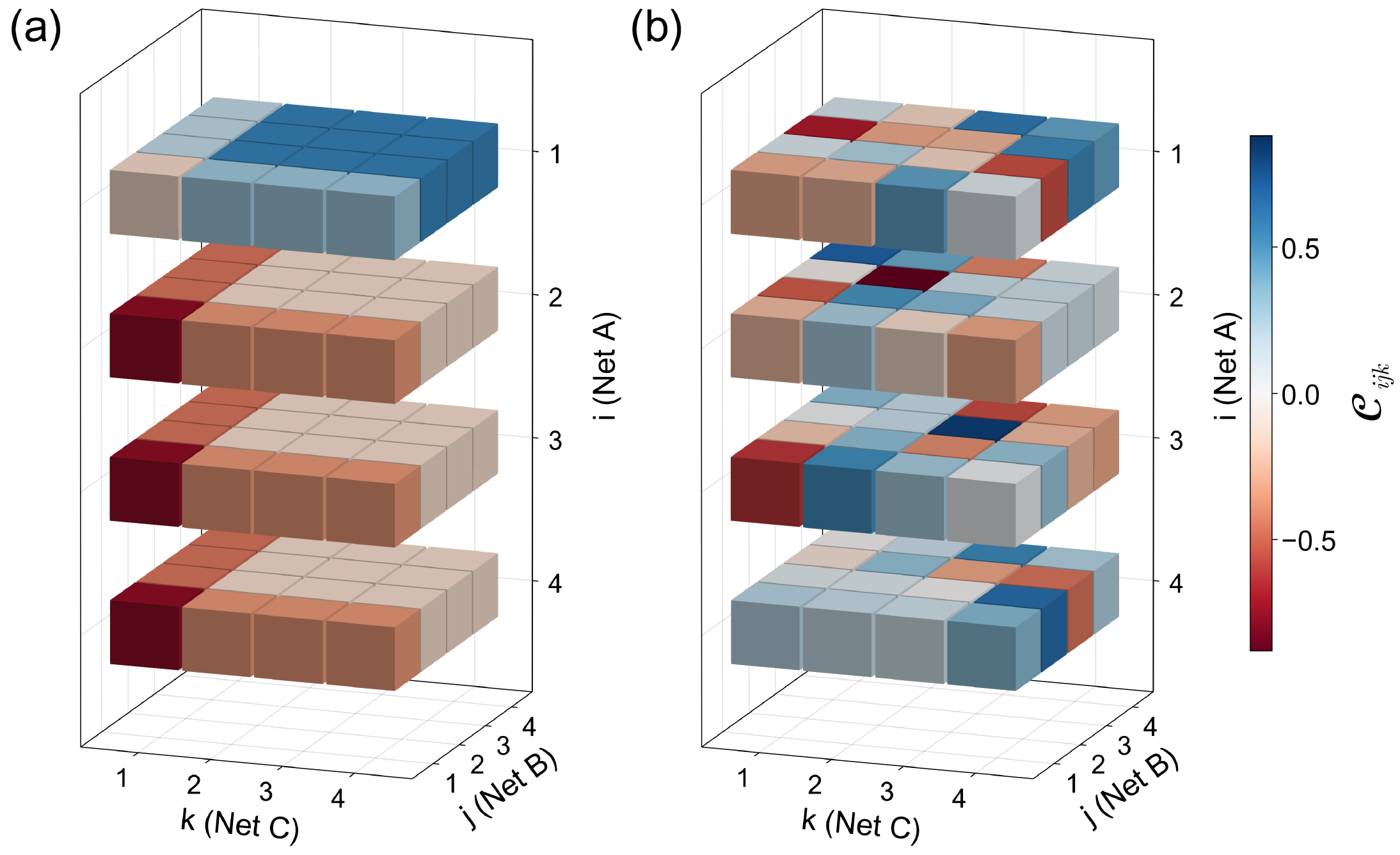}
	\caption{%
		Coupling tensors for the non-identical networks $K^A=K$, $K^B=0.9K$, and $K^C=1.1K$.
		(a) Optimal coupling tensor $\Ct_R^\star$ maximizing the rotational component and (b) random coupling tensor $\Ct_{R,\mathrm{rand}}$ with the same Frobenius norm $\norm{\Ct}_F=\kappa=3.0$.
		Each cube represents the element $\Ct_{ijk}$, where the indices $i$, $j$, and $k$ label the elements of networks $A$, $B$, and $C$, respectively, and its color represents the value of $\Ct_{ijk}$ on the common scale shown on the right.
	}
	\label{fig:tensor_R}
\end{figure}

We next consider the optimization of the rotational component, using the non-identical networks $K^A=K$, $K^B=0.9K$, and $K^C=1.1K$.
Following the procedure in section~\ref{sec:num_calc}, we adopt the result giving the largest $R$ among those whose fixed points are stable spirals.
The resulting optimal coupling tensor $\Ct_R^\star$ is shown in figure~\ref{fig:tensor_R}(a) and explicitly listed in Appendix~\ref{app:tensors}.
It gives the fixed point $(\phi_1^*,\phi_2^*)=(3.6740, 0.9007)$ with $R=0.7994$ and $\Lambda=1.3882$, i.e., the eigenvalues of $J$ are $-1.3882\pm0.7994i$, so that the fixed point is a stable spiral.
For comparison, the random coupling tensor $\Ct_{R,\mathrm{rand}}$ shown in figure~\ref{fig:tensor_R}(b) and explicitly listed in Appendix~\ref{app:tensors} gives the fixed point $(\phi_1^*,\phi_2^*)=(2.1990, 2.5461)$ with $R=0.0224$ and $\Lambda=0.2353$.
As discussed in section~\ref{sec:num_calc}, the locations of the fixed points have no special meaning for non-identical networks, and the values of $\Lambda$ and $R$ for the two couplings can be compared directly.

\begin{figure} [!t]
	\centering
	\includegraphics[width=\textwidth]{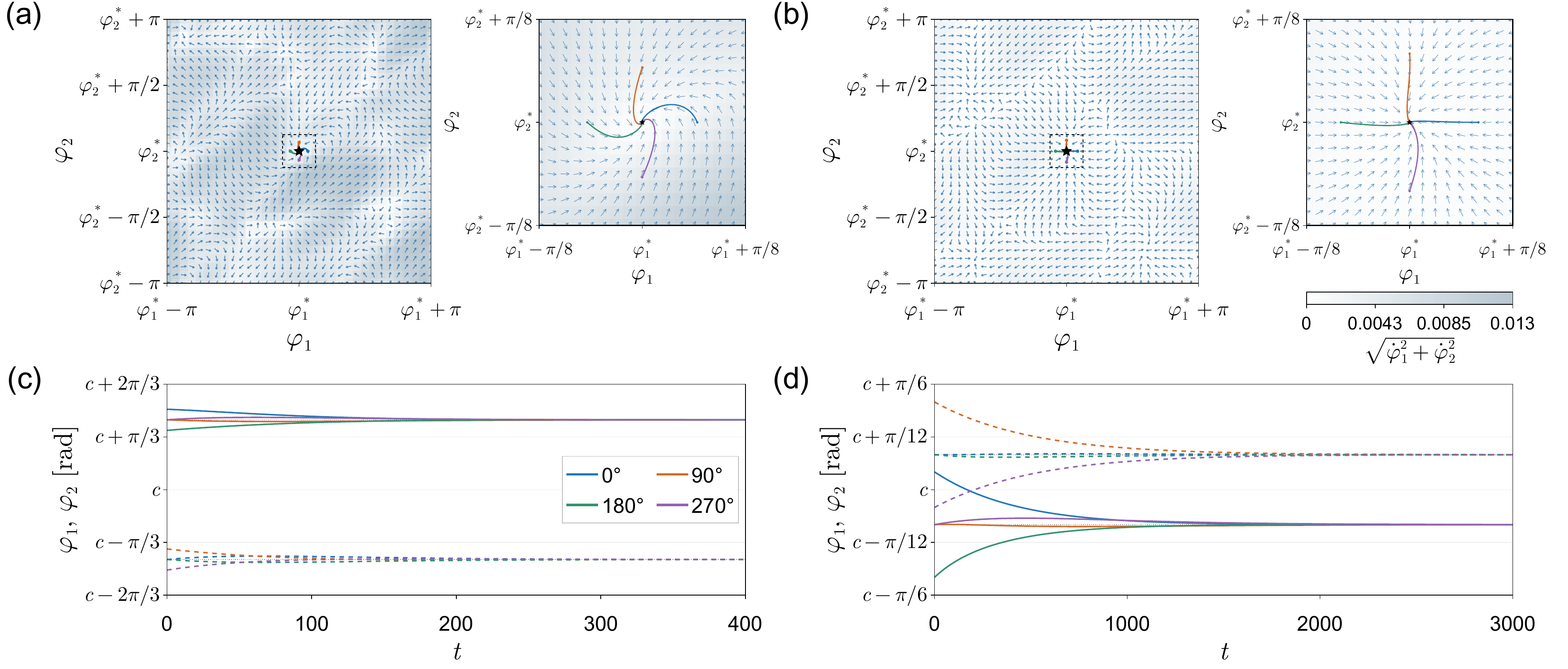}
	\caption{%
		Phase-difference dynamics for the non-identical networks.
		(a), (b) Vector fields of the phase-difference equations~\eqref{eq:phasediff_1} and \eqref{eq:phasediff_2} for the optimal coupling tensor $\Ct_R^\star$ and for the random coupling tensor $\Ct_{R,\mathrm{rand}}$ in figure~\ref{fig:tensor_R}, respectively.
		The arrows show the direction of the flow, the background color shows its speed $\sqrt{\dot{\phi}_1^2+\dot{\phi}_2^2}$ on a common scale, and the star indicates the stable fixed point $(\phi_1^*,\phi_2^*)$, whose location differs between the two couplings.
		The right panels of (a) and (b) magnify the dashed square around the fixed point, where the colored curves show the trajectories starting from the four initial points located at distances $\pi/15$ in (a) and $\pi/12$ in (b) from the fixed point in the directions $0^\circ$, $90^\circ$, $180^\circ$, and $270^\circ$ measured from the positive $\phi_1$ direction, as labeled in the legend in (c).
		The values of $(\phi_1^*,\phi_2^*)$, $\Lambda$, and $R$ are listed in table~\ref{tab:results}.
		(c), (d) Time evolution of the phase differences $\phi_1$ (solid lines) and $\phi_2$ (dashed lines) for the trajectories shown in the right panels of (a) and (b), respectively, where the vertical axes are labeled relative to $c=(\phi_1^*+\phi_2^*)/2$ and the horizontal dotted lines indicate $\phi_1^*$ and $\phi_2^*$.
	}
	\label{fig:opt_R}
\end{figure}

Figures~\ref{fig:opt_R}(a) and \ref{fig:opt_R}(b) show the vector fields of the phase-difference equations~\eqref{eq:phasediff_1} and \eqref{eq:phasediff_2} for the two couplings, together with trajectories starting at distances $\pi/15$ and $\pi/12$, respectively, from the fixed points in the four directions.
The background color represents the speed $\sqrt{\dot{\phi}_1^2+\dot{\phi}_2^2}$ of the phase differences on a common scale, and the right panels magnify the regions near the fixed points.
Figures~\ref{fig:opt_R}(c) and \ref{fig:opt_R}(d) show the time evolution of the phase differences along these trajectories.
The optimal coupling gives a rotational component about $35.7$ times larger than that of the random coupling, and the trajectories correspondingly approach the fixed point while rotating around it, whereas the trajectories for the random coupling approach the fixed point with little rotation.

\subsection{Validity of the optimal couplings}
\label{sec:discussion}
The results of both optimizations are summarized in
table~\ref{tab:results}.
The optimal coupling tensors derived in section~\ref{sec:optimization}
rely on conditions that are not guaranteed in advance and should be
verified for the coupling tensor actually obtained.
We summarize here which conditions are required for each optimization
and what happens when they fail.

For the linear stability, the decay rate $\Lambda$ characterizes the convergence to the phase-locked state only in the regime $4\tau\geq\sigma^2$, which the optimal coupling $\Ct_\Lambda^\star$ does not necessarily satisfy.
When $4\tau<\sigma^2$, the two eigenvalues of $J$ are real and distinct, and the actual convergence rate is given by the decay rate of the slower mode, which is smaller than $\Lambda$.
Therefore, as shown in Appendix~\ref{app:slow_mode}, $\Lambda^{\max}$ in equation~\eqref{eq:linear_fixed_norm_lst} is an upper bound on the actual convergence rate.
This bound is attained by $\Ct_\Lambda^\star$ if and only if the scalar $s=\bm{w}^TP\bm{w}$ is nonnegative, which is the case in section~\ref{sec:result_Lambda}, so that $\Ct_\Lambda^\star$ is then optimal also for the actual convergence rate.
When $s<0$, the optimization has to be reformulated, either by imposing $4\tau\geq\sigma^2$ as an additional constraint, as discussed in Appendix~\ref{app:kkt_formulation}, or by directly maximizing the decay rate of the slower mode, which requires a numerical optimization, as discussed in Appendix~\ref{app:slow_mode}.

For the rotational component, two conditions are required.
First, the matrix $P$ must possess a positive eigenvalue; otherwise,
$R=0$ for every coupling tensor.
Second, $\bm{w}^T\bm{v}_{\max}\neq0$ must hold, with the sign of
$\bm{v}_{\max}$ chosen such that $\bm{w}^T\bm{v}_{\max}>0$, for the
resulting phase-locked state to be linearly stable.
Both conditions are satisfied in section~\ref{sec:result_R}, but they
can fail, for example, when the networks possess symmetries.
For identical networks with common three-body coupling functions and
the coupling basis~\eqref{eq:H_basis}, the coupling tensor maximizing
$R$, whenever a positive $R$ is attainable, gives $\Lambda=0$ and thus
leaves the phase-locked state neutrally stable.
In the setting of section~\ref{sec:result_Lambda}, the
elements of each tree network are of only two kinds, the root and the
equivalent leaves, so that $R=0$ holds for every such coupling tensor
and no rotational motion can be generated at all, as shown in
Appendix~\ref{app:identical_symmetry}.
We therefore used the non-identical
networks, for which this is not the case, in section~\ref{sec:result_R}.
These situations are discussed in Appendix~\ref{app:rotation}.

\begin{table}[H]
	\centering
	\caption{%
		Fixed point $(\phi_1^*,\phi_2^*)$, linear decay rate $\Lambda$, and
		rotational component $R$ for the optimal and random coupling tensors.
		The values of $\Lambda$ and $R$ are evaluated at the respective stable
		fixed points.
		All coupling tensors have the same Frobenius norm
		$\norm{\Ct}_F=\kappa=3.0$.
	}
	\label{tab:results}
	\begin{tabular}{@{}llccc@{}}
		\toprule
		Networks & Coupling & $(\phi_1^*,\phi_2^*)$ & $\Lambda$ & $R$\\
		\midrule
		Identical & Optimal for $\Lambda$ & $(0,0)$ & 4.6220 & 0\\
		& Random & $(0,0)$ & 1.0342 & 0\\
		Non-identical & Optimal for $R$ & $(3.6740,0.9007)$ & 1.3882 & 0.7994\\
		& Random & $(2.1990,2.5461)$ & 0.2353 & 0.0224\\
		\bottomrule
	\end{tabular}
\end{table}

\section{Conclusion}
We investigated the optimal design of weak higher-order internetwork coupling among three collectively oscillating networks.
By extending the phase reduction framework for collective oscillations, we reduced the full network dynamics to a two-dimensional system of phase-difference equations.
We then formulated optimization problems for the coupling tensor under a constraint on its Frobenius norm and analytically derived the optimal tensors that maximize either the linear stability of the phase-locked state or the rotational property of the phase differences around it.
Numerical results for three networks, each consisting of four FitzHugh--Nagumo elements coupled in a tree structure, confirmed that the two optimized couplings yield, respectively, faster convergence to the phase-locked state and stronger rotational motion of the phase differences around it than a random coupling tensor with the same Frobenius norm.

Several extensions of the optimization problems are of interest.
One extension is to treat the linear stability and the rotational property as a single objective, for example, by maximizing a weighted sum of $\Lambda$ and $R$ or by maximizing one of them under a constraint on the other.
Another extension is to design the pairwise and three-body internetwork couplings together, since real systems generally involve both pairwise and higher-order interactions.

The present framework can also be extended in several other directions.
First, we measured the magnitude of the coupling by the Frobenius norm, which yields a dense coupling tensor.
In the pairwise case, the $L_1$ norm was also considered in Ref.~\cite{nakao2021sparse}, where it yielded sparse internetwork coupling matrices with only a small number of connections.
Since the linear decay rate $\Lambda$ is a linear function of the coupling vector, its maximization under an $L_1$-norm constraint can likewise be cast into a linear-programming problem.
This is practically important here because the number of hyperedges, $N^AN^BN^C$, grows as the product of the network sizes.
A sparse solution would identify the small number of element triplets that are essential for synchronization, thereby making the implementation of the designed coupling more efficient.

Second, the present framework is based on phase reduction, which is valid only when the internetwork coupling is sufficiently weak.
A phase-amplitude reduction framework for collectively oscillating networks has recently been developed~\cite{mircheski2023phase}, in which the deviation of the network state from the collective limit cycle is described by amplitude variables in addition to the collective phase.
Incorporating the amplitude dynamics would allow us to treat stronger internetwork coupling and to take into account the amplitude deviations it induces.

Third, the quantities required in the present framework, namely the collective limit cycles, the collective phase sensitivity functions, and the effective collective phase coupling functions, could be estimated from time-series data by some data-driven methods~\cite{brunton2019data}, without explicit knowledge of the governing equations.
This would extend the applicability of the present optimization method to experimental systems whose dynamical equations are unknown.

More broadly, since higher-order interactions can qualitatively alter collective dynamics, their optimal design is an important problem at the intersection of network science and control theory.
We expect that the present framework, which provides analytical solutions to the optimization problems for a basic three-network system, will serve as a starting point for further studies in this direction.

\clearpage

\section*{Acknowledgments}
We acknowledge JSPS KAKENHI JP22K14274 and JST PRESTO JPMJPR24K3 for financial support.







\appendix
\numberwithin{equation}{section}
\renewcommand{\theequation}{\Alph{section}.\arabic{equation}}

\section{Three identical networks with common three-body coupling functions}
\label{sec:appendix_identical}

In this appendix, as a special case of the general non-identical three-network system treated in the main text, we consider the case in which the three networks are identical.
That is, the numbers of elements are equal, $N^A=N^B=N^C\equiv N$, and the intrinsic dynamics $\bm{f}_i$, the intranetwork couplings $\bm{g}_{ij}$, and the prescribed coupling functions $\bm{H}_{ijk}$ are common to the three networks.
Consequently, the collective limit cycle $\{\bar{\bm{x}}_i(\theta)\}_{i=1}^{N}$, the collective frequency $\omega$ (so that $\Delta^A=\Delta^B=\Delta^C=0$), and the collective phase sensitivity functions $\bm{Q}_i(\theta)$ are also common to the three networks.

Throughout this appendix, we further assume that the coupling tensor is invariant under cyclic permutations of its indices,
\begin{equation}
	\Ct_{ijk}=\Ct_{jki}=\Ct_{kij},
	\label{eq:cyclic_invariance}
\end{equation}
so that the three-body coupling functions are also common to the three networks, as shown below.
We first show this property and derive the collective phase equations, then show that the in-phase state is a phase-locked state and analyze its linear stability, and finally optimize the coupling tensor within the class of cyclically invariant coupling tensors.

\subsection{Collective phase equations}
\label{app:identical_phase}

For identical networks, the three-body coupling functions in equation~\eqref{eq:decomposition}, written in terms of the local index order $(i,j,k)$ of each network, are given by
\begin{equation}
	\bm{h}_{ijk}^A=\Ct_{ijk}\bm{H}_{ijk},
	\qquad
	\bm{h}_{ijk}^B=\Ct_{kij}\bm{H}_{ijk},
	\qquad
	\bm{h}_{ijk}^C=\Ct_{jki}\bm{H}_{ijk},
	\label{eq:h_identical}
\end{equation}
which differ only in the order of the indices of the coupling tensor.
Under the cyclic invariance~\eqref{eq:cyclic_invariance}, these three functions therefore coincide, $\bm{h}_{ijk}^X=\bm{h}_{ijk}\equiv\Ct_{ijk}\bm{H}_{ijk}$ for $X\in\{A,B,C\}$; that is, the three networks are identical including their three-body coupling functions.
The full dynamical system is given by equations~\eqref{eq:model_A}--\eqref{eq:model_C} with $\bm{f}_i^X=\bm{f}_i$, $\bm{g}_{ij}^X=\bm{g}_{ij}$, $\bm{h}_{ijk}^X=\bm{h}_{ijk}$, and $N^X=N$ for $X\in\{A,B,C\}$.
Since the collective limit cycles and the collective phase sensitivity functions are common to the three networks, the individual phase coupling function in equation~\eqref{eq:gamma_elem} is also common and reduces to
\begin{equation}
	\gamma_{ijk}(\varphi,\varphi')
	=\frac{1}{2\pi}\int_0^{2\pi}
	\bm{Q}_i(\psi)\cdot
	\bm{h}_{ijk}\!\left(
	\bar{\bm{x}}_i(\psi),
	\bar{\bm{x}}_j(\psi+\varphi),
	\bar{\bm{x}}_k(\psi+\varphi')
	\right)\,\dd\psi,
	\label{eq:gamma_identical}
\end{equation}
and so does the effective collective phase coupling function in equation~\eqref{eq:Gamma},
\begin{equation}
	\Gamma(\varphi,\varphi')=\sum_{i=1}^{N}\sum_{j=1}^{N}\sum_{k=1}^{N}\gamma_{ijk}(\varphi,\varphi').
	\label{eq:Gamma_identical}
\end{equation}
The phase equations~\eqref{eq:phase_A}--\eqref{eq:phase_C} therefore reduce to
\begin{align}
	\dot{\theta}^A&=\omega+\epsilon\Gamma(\phi_{AB},\phi_{AC}),\\
	\dot{\theta}^B&=\omega+\epsilon\Gamma(\phi_{BC},\phi_{BA}),\\
	\dot{\theta}^C&=\omega+\epsilon\Gamma(\phi_{CA},\phi_{CB}),
\end{align}
where $\phi_{XY}=\theta^Y-\theta^X$ are the phase differences introduced in equation~\eqref{eq:phase_diff}.

\subsection{Phase-difference equations and linear stability analysis}
\label{app:identical_stability}

With $\Delta^B-\Delta^A=\Delta^C-\Delta^A=0$ and $\Gamma^A=\Gamma^B=\Gamma^C=\Gamma$, the phase-difference equations~\eqref{eq:phasediff_1} and \eqref{eq:phasediff_2} reduce to
\begin{align}
	\dot{\phi}_1&=\epsilon\left[\Gamma(\phi_2-\phi_1,-\phi_1)-\Gamma(\phi_1,\phi_2)\right],\label{eq:phasediff_identical_1}\\
	\dot{\phi}_2&=\epsilon\left[\Gamma(-\phi_2,\phi_1-\phi_2)-\Gamma(\phi_1,\phi_2)\right].\label{eq:phasediff_identical_2}
\end{align}
Since the right-hand sides vanish at $\phi_1=\phi_2=0$, the in-phase phase-locked state is a fixed point, and we take $(\phi_1^*,\phi_2^*)=(0,0)$ in what follows.
Evaluating the derivatives in equations~\eqref{eq:derivs_A}--\eqref{eq:derivs_C} at this fixed point, we find that the six quantities reduce to the two common values
\begin{equation}
	\Gamma_1=\partial_1\Gamma(0,0),
	\qquad
	\Gamma_2=\partial_2\Gamma(0,0),
	\label{eq:Gamma12_identical}
\end{equation}
i.e., $\Gamma_m^X=\Gamma_m$ for $m=1,2$ and $X\in\{A,B,C\}$.
The Jacobian matrix~\eqref{eq:J_general} then becomes
\begin{equation}
	J=
	\begin{pmatrix}
		-2\Gamma_1-\Gamma_2 & \Gamma_1-\Gamma_2\\
		\Gamma_2-\Gamma_1 & -\Gamma_1-2\Gamma_2
	\end{pmatrix},
	\label{eq:J_identical}
\end{equation}
where, as in the main text, the small parameter $\epsilon$ is excluded from $J$.
From equation~\eqref{eq:sigma_tau}, we have
\begin{equation}
	\sigma=3(\Gamma_1+\Gamma_2),
	\qquad
	\tau=3\left(\Gamma_1^2+\Gamma_1\Gamma_2+\Gamma_2^2\right),
	\label{eq:sigma_tau_identical}
\end{equation}
so that $4\tau-\sigma^2=3(\Gamma_1-\Gamma_2)^2\geq0$ holds for every coupling tensor considered here.
The condition $4\tau\geq\sigma^2$ assumed in the main text is thus automatically satisfied in this case, in contrast to the general non-identical case, where the eigenvalues can be real and distinct.
The eigenvalues~\eqref{eq:eigen} reduce to
\begin{equation}
	\lambda_{\pm}=-\frac{3}{2}(\Gamma_1+\Gamma_2)
	\pm i\frac{\sqrt{3}}{2}(\Gamma_1-\Gamma_2),
	\label{eq:eigen_identical}
\end{equation}
and the linear decay rate and the rotational component defined in equation~\eqref{eq:Lambda_R} become
\begin{equation}
	\Lambda=\frac{3}{2}(\Gamma_1+\Gamma_2),
	\qquad
	R=\frac{\sqrt{3}}{2}\left|\Gamma_1-\Gamma_2\right|.
	\label{eq:Lambda_R_identical}
\end{equation}
Hence, for $\Gamma_1\ne\Gamma_2$ the eigenvalues form the complex-conjugate pair $-\Lambda\pm iR$, and the fixed point is a stable spiral if $\Lambda>0$.
For $\Gamma_1=\Gamma_2$, the Jacobian matrix reduces to $J=-3\Gamma_1 I$, where $I$ is the identity matrix, and no rotational motion occurs.
These results hold for any coupling tensor satisfying equation~\eqref{eq:cyclic_invariance}, including both the optimal and the random coupling tensors used in section~\ref{sec:result_Lambda}.

\subsection{Optimal coupling tensors}
\label{app:identical_optimization}

Let $V_{ijk}$ be the function obtained by replacing $\bm{h}_{ijk}$ in equation~\eqref{eq:gamma_identical} with $\bm{H}_{ijk}$, so that $\gamma_{ijk}=\Ct_{ijk}V_{ijk}$.
Since $V_{ijk}^X=V_{ijk}$ for identical networks and all arguments of $V_{ijk}^X$ in equation~\eqref{eq:grad_vectors} vanish at $(0,0)$, the gradient vectors at this fixed point are given by
\begin{equation}
	(\bm{a}_m^A)_{\ell(i,j,k)}=\partial_mV_{ijk}(0,0),
	\qquad
	(\bm{a}_m^B)_{\ell(i,j,k)}=\partial_mV_{jki}(0,0),
	\qquad
	(\bm{a}_m^C)_{\ell(i,j,k)}=\partial_mV_{kij}(0,0),
	\label{eq:grad_identical}
\end{equation}
for $m=1,2$.
In addition to the vector $\bm{w}$ in equation~\eqref{eq:w}, we define
\begin{equation}
	\bm{z}=\sum_{X\in\{A,B,C\}}\left(\bm{a}_1^{X}-\bm{a}_2^{X}\right).
	\label{eq:z}
\end{equation}
From equation~\eqref{eq:grad_identical}, the components of $\bm{w}$ and $\bm{z}$ are given by
\begin{equation}
	\begin{aligned}
		(\bm{w})_{\ell(i,j,k)}&=\sum_{(p,q,r)}\left[\partial_1V_{pqr}(0,0)+\partial_2V_{pqr}(0,0)\right],\\
		(\bm{z})_{\ell(i,j,k)}&=\sum_{(p,q,r)}\left[\partial_1V_{pqr}(0,0)-\partial_2V_{pqr}(0,0)\right],
	\end{aligned}
	\label{eq:wz_components}
\end{equation}
where the sums are taken over the three cyclic permutations $(p,q,r)=(i,j,k)$, $(j,k,i)$, and $(k,i,j)$.
Hence, both $\bm{w}$ and $\bm{z}$ are invariant under cyclic permutations of $(i,j,k)$.

Since $\Gamma_m^X=\Gamma_m$ for all $X$, equation~\eqref{eq:Gamma_linear} gives
\begin{equation}
	\begin{aligned}
		\Gamma_1+\Gamma_2&=\frac{1}{3}\sum_{X\in\{A,B,C\}}\left(\Gamma_1^X+\Gamma_2^X\right)=\frac{1}{3}\bm{w}^T\bm{c},\\
		\Gamma_1-\Gamma_2&=\frac{1}{3}\sum_{X\in\{A,B,C\}}\left(\Gamma_1^X-\Gamma_2^X\right)=\frac{1}{3}\bm{z}^T\bm{c}.
	\end{aligned}
\end{equation}
Substituting these into equation~\eqref{eq:Lambda_R_identical}, we express the linear decay rate and the rotational component as
\begin{equation}
	\Lambda(\bm{c})=\frac{1}{2}\bm{w}^T\bm{c},
	\qquad
	R(\bm{c})=\frac{\sqrt{3}}{6}\left|\bm{z}^T\bm{c}\right|.
	\label{eq:Lambda_R_wz}
\end{equation}
By the Cauchy--Schwarz inequality, $\Lambda$ and $R$ under the norm constraint~\eqref{eq:norm_constraint} are maximized by $\bm{c}$ in the direction of $\bm{w}$ and $\pm\bm{z}$, respectively.
For $\bm{w}\neq\bm{0}$, the optimal coupling vector maximizing $\Lambda$ and the corresponding maximum value are
\begin{equation}
	\bm{c}_{\Lambda}^{\star}
	=\kappa\,\frac{\bm{w}}{\norm{\bm{w}}_2},
	\qquad
	\Lambda^{\max}=\frac{\kappa}{2}\norm{\bm{w}}_2,
	\label{eq:opt_Lambda_identical}
\end{equation}
and, for $\bm{z}\neq\bm{0}$, those maximizing $R$ are
\begin{equation}
	\bm{c}_{R}^{\star}
	=\pm\kappa\,\frac{\bm{z}}{\norm{\bm{z}}_2},
	\qquad
	R^{\max}=\frac{\sqrt{3}\kappa}{6}\norm{\bm{z}}_2,
	\label{eq:opt_R_identical}
\end{equation}
where the sign is fixed by requiring $\Lambda(\bm{c}_R^{\star})\geq0$ as in the main text.
If $\bm{w}=\bm{0}$ or $\bm{z}=\bm{0}$, then $\Lambda(\bm{c})=0$ or $R(\bm{c})=0$, respectively, for all cyclically invariant $\bm{c}$.

Since $\bm{w}$ and $\bm{z}$ are invariant under cyclic permutations, so are these optimal coupling vectors; that is, they satisfy the assumption~\eqref{eq:cyclic_invariance}, and $(0,0)$ is indeed a fixed point for them.

\subsection{Coupling functions symmetric in the two source elements}
\label{app:identical_symmetry}

We finally consider a coupling function that is symmetric with respect to its two source elements, $\bm{H}_{ijk}=\bm{H}_{ikj}$, such as the sinusoidal basis~\eqref{eq:H_basis} used in the numerical example.
Then equation~\eqref{eq:gamma_identical} implies $V_{ijk}(\varphi,\varphi')=V_{ikj}(\varphi',\varphi)$, and differentiating this relation at $(0,0)$ yields
\begin{equation}
	\partial_1V_{ijk}(0,0)=\partial_2V_{ikj}(0,0).
	\label{eq:swap_relation}
\end{equation}
Since exchanging the last two indices maps the three cyclic permutations of $(i,j,k)$ to those of $(i,k,j)$ and, by equation~\eqref{eq:swap_relation}, interchanges $\partial_1V$ and $\partial_2V$, equation~\eqref{eq:wz_components} gives
\begin{equation}
	(\bm{w})_{\ell(i,k,j)}=(\bm{w})_{\ell(i,j,k)},
	\qquad
	(\bm{z})_{\ell(i,k,j)}=-(\bm{z})_{\ell(i,j,k)}.
	\label{eq:wz_swap}
\end{equation}
Relabeling $j$ and $k$ in the sum $\bm{w}^T\bm{z}=\sum_{i,j,k}(\bm{w})_{\ell(i,j,k)}(\bm{z})_{\ell(i,j,k)}$ then yields $\bm{w}^T\bm{z}=-\bm{w}^T\bm{z}$, i.e.,
\begin{equation}
	\bm{w}^T\bm{z}=0.
	\label{eq:orthogonality_identical}
\end{equation}

From equations~\eqref{eq:Lambda_R_wz}--\eqref{eq:opt_R_identical}, this gives $R(\bm{c}_\Lambda^{\star})=0$ and, when $\bm{z}\neq\bm{0}$, $\Lambda(\bm{c}_R^{\star})=0$.
That is, the optimal coupling for the linear stability generates no rotation, consistently with the non-oscillatory approach observed in section~\ref{sec:result_Lambda}, whereas the optimal coupling for the rotational component would leave the phase-locked state neutrally stable.
Accordingly, since $\bm{c}^TP\bm{c}=4\tau-\sigma^2=(\bm{z}^T\bm{c})^2/3$ for cyclically invariant $\bm{c}$, the scalar $s=\bm{w}^TP\bm{w}$ introduced in Appendix~\ref{app:kkt} vanishes.

In the numerical example, however, $\bm{z}=0$ identically.
To see this, we write $D_{pqr}=\partial_1V_{pqr}(0,0)-\partial_2V_{pqr}(0,0)$, so that equation~\eqref{eq:wz_components} gives $(\bm{z})_{\ell(i,j,k)}=D_{ijk}+D_{jki}+D_{kij}$, and equation~\eqref{eq:swap_relation} gives
\begin{equation}
	D_{pqr}=\partial_2V_{prq}(0,0)-\partial_2V_{pqr}(0,0).
	\label{eq:D_swap}
\end{equation}
The elements of each tree network are of only two kinds, the root and the equivalent leaves, so that $V_{pqr}$ depends on $p$, $q$, and $r$ only through their kinds.
Since every triple $(i,j,k)$ contains at least two elements of the same kind, its three cyclic permutations have the kinds $(a,b,b)$, $(b,b,a)$, and $(b,a,b)$ for some kinds $a$ and $b$, where $a=b$ is also allowed.
Denoting the kinds by the subscripts, equation~\eqref{eq:D_swap} then gives
\begin{equation}
	D_{abb}=0,
	\qquad
	D_{bba}=\partial_2V_{bab}(0,0)-\partial_2V_{bba}(0,0)=-D_{bab},
\end{equation}
so that $(\bm{z})_{\ell(i,j,k)}=D_{abb}+D_{bba}+D_{bab}=0$ for every $(i,j,k)$, i.e., $\bm{z}=\bm{0}$.
Equation~\eqref{eq:Lambda_R_wz} then gives $R=0$ for every cyclically invariant coupling tensor, which means that no rotational motion can be generated at all.
This cancellation relies on the three networks being identical.
For the non-identical networks in section~\ref{sec:result_R}, the functions $V_{ijk}^X$ differ among the networks, and a positive rotational component is indeed obtained.

\section{Optimization of the linear stability beyond the regime $4\tau\geq\sigma^2$}
\label{app:kkt}

In the main text, we restricted our attention to the case $4\tau\geq\sigma^2$, i.e., $\bm{c}^TP\bm{c}\geq0$, but the optimal coupling vector $\bm{c}_\Lambda^\star$ in equation~\eqref{eq:linear_fixed_norm} was obtained without imposing this condition.
In this appendix, we discuss how the optimization should be formulated when this condition is not automatically satisfied.

\subsection{Constrained formulation}
\label{app:kkt_formulation}

Imposing the condition explicitly as a constraint, we consider
\begin{equation}
	\begin{aligned}
		\max_{\bm{c}}\quad & \bm{w}^T\bm{c}\\
		\text{subject to}\quad & \norm{\bm{c}}_2=\kappa,
		\quad \bm{c}^TP\bm{c}\geq0,
	\end{aligned}
	\label{eq:opt_Lambda_constrained}
\end{equation}
which is a constrained optimization problem with an inequality constraint and can be treated by the Karush--Kuhn--Tucker conditions.
With multipliers $\mu$ and $\nu\geq0$ for the two constraints, the stationarity condition of the Lagrangian $\bm{w}^T\bm{c}-\mu(\norm{\bm{c}}_2^2-\kappa^2)+\nu\,\bm{c}^TP\bm{c}$ gives
\begin{equation}
	\bm{c}(t)\propto(I-tP)^{-1}\bm{w},
	\qquad t=\nu/\mu,
	\label{eq:kkt_stationarity}
\end{equation}
where $I$ is the identity matrix and the overall scale is fixed by the norm constraint.
Since the complementary slackness condition $\nu\,\bm{c}^TP\bm{c}=0$ requires $\nu=0$ whenever the inequality constraint is inactive, the candidate solutions are separated into two cases.

When $\nu=0$, the inequality constraint does not enter the stationarity condition, and equation~\eqref{eq:kkt_stationarity} reduces to
$\bm{c}\propto\bm{w}$, i.e., the solution $\bm{c}_\Lambda^\star$ of the main text.
Since $(\bm{c}_\Lambda^\star)^TP\bm{c}_\Lambda^\star=\kappa^2 s/\norm{\bm{w}}_2^2$ with $s=\bm{w}^TP\bm{w}$, this case is admissible if and only if $s\geq0$, so the sign of the single scalar $s$ tells in advance whether the analytical solution is consistent with the assumed regime.
In the numerical example in section~\ref{sec:result_Lambda}, $s=0$, so that this case applies with the inequality constraint satisfied as an equality.

When $\nu>0$, the constraint is active and the solution lies on the boundary $\bm{c}^TP\bm{c}=0$, where the two eigenvalues of $J$ degenerate into the double eigenvalue $-\Lambda$.
The ratio $t$ is then determined by
the scalar equation $\bm{c}(t)^TP\bm{c}(t)=0$, a one-dimensional root-finding problem along the curve~\eqref{eq:kkt_stationarity} that is readily solved numerically.

\subsection{Optimization of the actual convergence rate}
\label{app:slow_mode}

The formulation~\eqref{eq:opt_Lambda_constrained} keeps the analysis within the regime in which $\Lambda$ characterizes the convergence, but it is not the optimization of the convergence rate itself.
For a general coupling vector, the decay rate of the slower mode, i.e., the actual local convergence rate toward the phase-locked state, is
\begin{equation}
	\Lambda_{\mathrm{eff}}(\bm{c})
	=-\max_{\pm}\mathrm{Re}\,\lambda_{\pm}
	=\frac{1}{2}\left[
	\bm{w}^T\bm{c}-\sqrt{\max\{-\bm{c}^TP\bm{c},0\}}
	\right],
	\label{eq:Lambda_eff}
\end{equation}
which coincides with $\Lambda(\bm{c})$ when $\bm{c}^TP\bm{c}\geq0$ and is strictly smaller otherwise.
Hence $\Lambda_{\mathrm{eff}}(\bm{c})\leq\Lambda(\bm{c})$ for every $\bm{c}$, and the problem~\eqref{eq:opt_Lambda} of the main text is a relaxation of the maximization of $\Lambda_{\mathrm{eff}}$.
Its optimal value $\Lambda^{\max}$ in equation~\eqref{eq:linear_fixed_norm_lst} is therefore an upper bound on the attainable convergence rate.
When $s\geq0$, the relaxation is tight, since $\bm{c}_\Lambda^\star$ then attains $\Lambda_{\mathrm{eff}}(\bm{c}_\Lambda^\star)=\Lambda^{\max}$ and is therefore globally optimal for the actual convergence rate as well.
When $s<0$, in contrast, $\bm{c}_\Lambda^\star$ lies in the region $\bm{c}^TP\bm{c}<0$ and gives $\Lambda_{\mathrm{eff}}(\bm{c}_\Lambda^\star)<\Lambda^{\max}$, so that it is in general not optimal for the actual convergence rate.
Since $\Lambda_{\mathrm{eff}}$ is nonlinear in $\bm{c}$ and not differentiable on the boundary $\bm{c}^TP\bm{c}=0$, its maximization in this case has to be performed numerically, and we leave it for future work.

\section{Cases where the rotational optimization does not work}
\label{app:rotation}

The optimization of the linear stability requires $\bm{w}\neq\bm{0}$, which is satisfied in general.
The optimization of the rotational component, in contrast, can fail as a consequence of the symmetry of the networks, and there are two situations in which it does not yield a useful coupling tensor.

In the first situation, a positive $R$ is attainable and the optimal coupling tensor is obtained, but the resulting phase-locked state is not linearly stable.
This occurs when $\bm{w}$ is orthogonal to $\bm{v}_{\max}$, in which case $\Lambda(\bm{c}_R^\star)=\kappa\bm{w}^T\bm{v}_{\max}/2=0$ and the fixed point is neutrally stable.

In the second situation, $R=0$ for every coupling tensor, so that no rotational motion can be generated at all.
This occurs when $P$ has no positive eigenvalue, i.e., $\nu_{\max}\leq0$, so that $\bm{c}^TP\bm{c}\leq0$ for every $\bm{c}$.

Both situations arise for identical networks with common three-body coupling functions and a coupling basis symmetric in the two source elements, as shown in Appendix~\ref{app:identical_symmetry}.
In this case, $\Lambda$ and $R$ are given by equation~\eqref{eq:Lambda_R_wz}, where $\bm{w}$ and $\bm{z}$ are orthogonal to each other by equation~\eqref{eq:orthogonality_identical}.
When $\bm{z}\neq\bm{0}$, the maximizer of $R$, $\bm{c}_R^\star\propto\bm{z}$, is orthogonal to $\bm{w}$ and gives $\Lambda=0$, which corresponds to the first situation.
When $\bm{z}=\bm{0}$, $R=0$ for every coupling tensor giving common three-body coupling functions, which corresponds to the second situation.
The latter occurs when the elements of each network are of at most two kinds, as in the tree networks in section~\ref{sec:result_Lambda}, and in particular when all elements are equivalent on the collective limit cycle.
In this case, the cyclic direction $A\to B\to C$ cannot be distinguished from its reverse.

\section{Coupling tensors used in the numerical example}
\label{app:tensors}

This appendix gives the numerical values of the coupling tensors shown in figures~\ref{fig:tensor_Lambda} and \ref{fig:tensor_R}.
All of them satisfy the norm constraint $\norm{\Ct}_F=\kappa=3.0$.
Each tensor is given as four matrices $(\Ct_{ijk})_{jk}$ with the index $i$ of network $A$ fixed, where the row and column indices are $j$ and $k$, referring to the elements of networks $B$ and $C$, respectively.
\begingroup
\small
\setlength{\arraycolsep}{3pt}
\subsection{Identical networks: optimal coupling for the linear stability}
\begin{equation}
	\Ct_{1jk}=\begin{pmatrix}
		0.6265 & 0.4961 & 0.4961 & 0.4961 \\
		0.4961 & 0.3831 & 0.3831 & 0.3831 \\
		0.4961 & 0.3831 & 0.3831 & 0.3831 \\
		0.4961 & 0.3831 & 0.3831 & 0.3831
	\end{pmatrix},
	\quad
	\Ct_{2jk}=\begin{pmatrix}
		0.4961 & 0.3831 & 0.3831 & 0.3831 \\
		0.3831 & 0.3000 & 0.3000 & 0.3000 \\
		0.3831 & 0.3000 & 0.3000 & 0.3000 \\
		0.3831 & 0.3000 & 0.3000 & 0.3000
	\end{pmatrix},
\end{equation}
\begin{equation}
	\Ct_{3jk}=\begin{pmatrix}
		0.4961 & 0.3831 & 0.3831 & 0.3831 \\
		0.3831 & 0.3000 & 0.3000 & 0.3000 \\
		0.3831 & 0.3000 & 0.3000 & 0.3000 \\
		0.3831 & 0.3000 & 0.3000 & 0.3000
	\end{pmatrix},
	\quad
	\Ct_{4jk}=\begin{pmatrix}
		0.4961 & 0.3831 & 0.3831 & 0.3831 \\
		0.3831 & 0.3000 & 0.3000 & 0.3000 \\
		0.3831 & 0.3000 & 0.3000 & 0.3000 \\
		0.3831 & 0.3000 & 0.3000 & 0.3000
	\end{pmatrix}.
\end{equation}

\subsection{Identical networks: random coupling}
\begin{equation}
	\Ct_{1jk}=\begin{pmatrix}
		-0.0113 & 0.5101 & -0.1920 & 0.1327 \\
		0.5101 & -0.1010 & 0.0364 & 0.1664 \\
		-0.1920 & 0.1091 & 0.1762 & -0.5169 \\
		0.1327 & 0.1028 & 0.2130 & 0.6856
	\end{pmatrix},
	\quad
	\Ct_{2jk}=\begin{pmatrix}
		0.5101 & -0.1010 & 0.1091 & 0.1028 \\
		-0.1010 & 0.3818 & 0.5562 & 0.2602 \\
		0.0364 & 0.5562 & 0.4614 & -0.6078 \\
		0.1664 & 0.2602 & -0.2572 & 0.4285
	\end{pmatrix},
\end{equation}
\begin{equation}
	\Ct_{3jk}=\begin{pmatrix}
		-0.1920 & 0.0364 & 0.1762 & 0.2130 \\
		0.1091 & 0.5562 & 0.4614 & -0.2572 \\
		0.1762 & 0.4614 & 0.1540 & 0.1224 \\
		-0.5169 & -0.6078 & 0.1224 & -0.7257
	\end{pmatrix},
	\quad
	\Ct_{4jk}=\begin{pmatrix}
		0.1327 & 0.1664 & -0.5169 & 0.6856 \\
		0.1028 & 0.2602 & -0.6078 & 0.4285 \\
		0.2130 & -0.2572 & 0.1224 & -0.7257 \\
		0.6856 & 0.4285 & -0.7257 & -0.0888
	\end{pmatrix}.
\end{equation}

\subsection{Non-identical networks: optimal coupling for the rotational component}
\begin{equation}
	\Ct_{1jk}=\begin{pmatrix}
		-0.1432 & 0.3020 & 0.3020 & 0.3020 \\
		0.2066 & 0.5825 & 0.5825 & 0.5825 \\
		0.2066 & 0.5825 & 0.5825 & 0.5825 \\
		0.2066 & 0.5825 & 0.5825 & 0.5825
	\end{pmatrix},
	\quad
	\Ct_{2jk}=\begin{pmatrix}
		-0.7777 & -0.3740 & -0.3740 & -0.3740 \\
		-0.4680 & -0.1332 & -0.1332 & -0.1332 \\
		-0.4680 & -0.1332 & -0.1332 & -0.1332 \\
		-0.4680 & -0.1332 & -0.1332 & -0.1332
	\end{pmatrix},
\end{equation}
\begin{equation}
	\Ct_{3jk}=\begin{pmatrix}
		-0.7777 & -0.3740 & -0.3740 & -0.3740 \\
		-0.4680 & -0.1332 & -0.1332 & -0.1332 \\
		-0.4680 & -0.1332 & -0.1332 & -0.1332 \\
		-0.4680 & -0.1332 & -0.1332 & -0.1332
	\end{pmatrix},
	\quad
	\Ct_{4jk}=\begin{pmatrix}
		-0.7777 & -0.3740 & -0.3740 & -0.3740 \\
		-0.4680 & -0.1332 & -0.1332 & -0.1332 \\
		-0.4680 & -0.1332 & -0.1332 & -0.1332 \\
		-0.4680 & -0.1332 & -0.1332 & -0.1332
	\end{pmatrix}.
\end{equation}

\subsection{Non-identical networks: random coupling}
\begin{equation}
	\Ct_{1jk}=\begin{pmatrix}
		-0.3066 & -0.2744 & 0.4509 & 0.0757 \\
		0.0889 & 0.2406 & -0.1562 & -0.5585 \\
		-0.6946 & -0.3156 & -0.2759 & 0.5501 \\
		0.1216 & -0.1632 & 0.6054 & 0.4429
	\end{pmatrix},
	\quad
	\Ct_{2jk}=\begin{pmatrix}
		-0.2540 & 0.2635 & -0.1335 & -0.3222 \\
		-0.5294 & 0.5010 & 0.3686 & 0.1342 \\
		-0.0314 & -0.8809 & 0.1666 & 0.1550 \\
		0.7045 & 0.4274 & -0.4175 & 0.0944
	\end{pmatrix},
\end{equation}
\begin{equation}
	\Ct_{3jk}=\begin{pmatrix}
		-0.6263 & 0.5252 & 0.2801 & 0.0266 \\
		-0.2153 & 0.3428 & -0.3976 & 0.3107 \\
		0.0187 & 0.1820 & 0.8247 & -0.2578 \\
		0.3445 & 0.1800 & -0.5674 & -0.3147
	\end{pmatrix},
	\quad
	\Ct_{4jk}=\begin{pmatrix}
		0.2277 & 0.1819 & 0.1457 & 0.3632 \\
		0.0881 & 0.0907 & -0.0053 & 0.6553 \\
		-0.0982 & 0.3223 & -0.3251 & -0.4640 \\
		-0.0046 & 0.1741 & 0.5445 & 0.2456
	\end{pmatrix}.
\end{equation}
\endgroup
\end{document}